\pdfoutput=1
\documentclass[letterpaper,twocolumn,10pt]{article}
\usepackage{usenix-2020-09}
\PassOptionsToPackage{hyphens}{url}
\usepackage[english]{babel}

\usepackage{tikz}
\usetikzlibrary{arrows.meta,positioning,fit,backgrounds,calc,decorations.pathreplacing}
\usetikzlibrary{shapes.misc}
\newcommand{\figfontS}{\fontfamily{Crlt-LF}\fontsize{8}{9.2}\selectfont}
\newcommand{\kbn}[1]{{\usefont{U}{pzd}{m}{n}\char\numexpr181+#1\relax}}   %
\definecolor{kBlue}{HTML}{4472C4}   \definecolor{kBlueS}{HTML}{2F528F}   \definecolor{kBlueL}{HTML}{DAE3F3}   \definecolor{kBlueM}{HTML}{B4C7E7}
\definecolor{kOrange}{HTML}{ED7D31} \definecolor{kOrangeS}{HTML}{AE5A21} \definecolor{kOrangeL}{HTML}{FBE5D6} \definecolor{kOrangeM}{HTML}{F8CBAD}
\definecolor{kGray}{HTML}{A5A5A5}   \definecolor{kGrayS}{HTML}{787878}   \definecolor{kGrayL}{HTML}{EDEDED}   \definecolor{kGrayM}{HTML}{DBDBDB}
\definecolor{kGrayD}{HTML}{7F7F7F}
\definecolor{kGold}{HTML}{FFC000}   \definecolor{kGoldS}{HTML}{BC8C00}   \definecolor{kGoldL}{HTML}{FFF2CC}   \definecolor{kGoldM}{HTML}{FFE699}
\definecolor{kSky}{HTML}{5B9BD5}    \definecolor{kSkyS}{HTML}{41719C}    \definecolor{kSkyL}{HTML}{DEEBF7}    \definecolor{kSkyM}{HTML}{BDD7EE}
\definecolor{kGreen}{HTML}{70AD47}  \definecolor{kGreenS}{HTML}{507E32}  \definecolor{kGreenL}{HTML}{E2F0D9}  \definecolor{kGreenM}{HTML}{C5E0B4}
\definecolor{kRed}{HTML}{C0504D}    \definecolor{kRedS}{HTML}{943634}    %
\tikzset{
  kfig/.style={font=\figfontS, x=1cm, y=1cm},
  kregion/.style 2 args={draw=#1, fill=#2, line width=0.75pt},
  kpart/.style 2 args={draw=#1, fill=#2!50!white, line width=0.75pt, font=\figfontS, text=black, inner sep=1.2pt,
                       align=center},
  kcard/.style 2 args={draw=#1, fill=#2, line width=0.6pt, font=\figfontS, text=black, inner sep=1.2pt,
                       align=center},
  khdr/.style={font=\figfontS\bfseries, inner sep=0.8pt, text=black},
  ktxt/.style={font=\figfontS, inner sep=0.8pt, align=center, text=black},
  karrow/.style={-{Triangle[length=1.8mm,width=1.6mm]}, draw=#1, line width=1pt},
  karr/.style={karrow=kGrayD},
  kcell/.style={draw=kGrayS, line width=0.45pt, fill=white, minimum width=2.1mm, minimum height=2.6mm, inner sep=0pt},
  khit/.style={draw=kGreenS, line width=0.55pt, fill=kGreen!55!white, minimum width=2.1mm, minimum height=2.6mm, inner sep=0pt},
  kdash/.style={dash pattern=on 2.4pt off 1.4pt},
  kbrace/.style={decorate, decoration={brace, amplitude=1.2mm}, draw=black, line width=0.6pt},
  kcodep/.style={kcell, minimum width=2.3mm, minimum height=3.1mm, font=\figfontS},   %
  kcodeh/.style={khit, minimum width=2.3mm, minimum height=3.1mm, font=\figfontS},    %
  kpartn/.style={draw=kBlue, fill=kBlueL, line width=0.6pt},                          %
  kpartx/.style={draw=kGrayS, fill=white, line width=0.6pt, kdash},                   %
  ksel/.style={draw=kGoldS, line width=1.3pt},                                        %
  kar/.style={-{Triangle[length=1.2mm,width=1.1mm]}, draw=#1, line width=0.7pt}, kar/.default=kGrayD,  %
  kround/.style={rounded corners=3pt},                                                %
  kpill/.style={rounded corners=2.2mm},                                               %
}
\newcommand{\kstage}[9]{%
  \draw[kregion={#5}{#7}, rounded corners=3pt] (#1,#2) rectangle ++(#3,#4);
  \draw[kpart={#6}{#5}] (#1,#2+#4-0.46) [rounded corners=3pt] -- (#1,#2+#4) -- (#1+#3,#2+#4) [sharp corners] -- (#1+#3,#2+#4-0.46) -- cycle;
  \node[anchor=west, text=black, font=\figfontS\bfseries, inner sep=0.8pt] at (#1+0.12,#2+#4-0.23)
       {\kbn{#8}\hspace{2.5pt}#9};
}
\tikzset{
  kgm/.style={circle, minimum size=3.4mm, inner sep=0pt, draw=kGreenS, fill=kGreen!55!white, line width=0.6pt},
  kgx/.style={circle, minimum size=3.4mm, inner sep=0pt, draw=kGrayS, fill=kGrayL, line width=0.6pt, kdash},
  kge/.style={draw=black, line width=0.75pt},
  kgd/.style={draw=kGrayS, line width=0.6pt, kdash},
  kwait/.style={kregion={kSkyS}{kSkyL}, kdash}, kcomp/.style={kpart={kGreenS}{kGreen}},
  kpre/.style={kar=kSkyS, line width=0.8pt, kdash},
  kuse/.style={kar=kGreenS, line width=0.8pt},
  kpos/.style={kcell, fill=kSkyM, draw=kSkyS},
}
\usepackage{xcolor,xspace}
\newif\ifrevmark \revmarkfalse

\usepackage{url}
\usepackage{placeins} %
\usepackage{xurl}    %
\usepackage{epsfig,graphicx,subfig,amsmath,amssymb}
\usepackage{multirow}
\usepackage{enumitem}
\usepackage[T1]{fontenc}
\usepackage{booktabs}
\usepackage{colortbl}
\usepackage{algorithm}
\usepackage[noend]{algpseudocode}
\usepackage{pifont}

\newcommand{\cmark}{\textcolor{green!60!black}{\ding{51}}}
\newcommand{\xmark}{\textcolor{red!75!black}{\ding{55}}}

\usepackage{balance}
\usetikzlibrary{patterns}

\newcommand{\sys}{KuroANN\xspace}

\newcommand{\pipeann}{PipeANN\xspace}
\newcommand{\pf}{PipeANN-Filter\xspace}
\newcommand{\pfpost}{PipeANN-Post\xspace}
\newcommand{\fdiskann}{Filtered-DiskANN\xspace}

\newcommand{\curator}{Curator\xspace}
\newcommand{\faiss}{Faiss\xspace}

\makeatletter
\def\section{\@startsection {section}{1}{\z@}{-2.4ex plus-1ex minus-.2ex}{1.5ex plus.2ex}{\reset@font\large\bf}}
\def\subsection{\@startsection{subsection}{2}{\z@}{-2.2ex plus-1ex minus-.2ex}{1.0ex plus.2ex}{\reset@font\normalsize\bf}}
\makeatother

\begin{document}

\title{\Large \bf \sys: Avoiding Unnecessary SSD Reads in High-Recall Filtered Vector Search}
\date{}
\author{
{\rm Soobin Cho$^{1}$, Nakyung Lee$^{1}$, Jiwoong Park$^{2}$, Gyuyeong Kim$^{1}$}\\
$^{1}$Yonsei University \quad $^{2}$Sungshin Women's University
}
\maketitle

\begin{abstract}
We present \sys{}, a new on-SSD filtered vector search system for partition indexes.
Existing solutions assign the same number of SSD reads to all queries within a given selectivity range, wasting I/O for many queries.
\sys{} instead assigns each query's SSD reads based on the number of vectors that pass its filter, called its \emph{match count}.
This is based on our observation that the match count is known before the search and that the number of SSD reads a query requires depends on this match count.
\sys{} measures offline how many partitions to probe and candidates to read for the target recall at each count.
At runtime, a query looks up these numbers, filters only the probed partitions, and stops reading from the SSD when further reads are unlikely to help.
This significantly avoids unnecessary SSD reads and filtering work, improving throughput and latency at high recall.
Experimental results show that \sys{} outperforms PipeANN-Filter and Faiss by up to 34.5$\times$ and 9.10$\times$, respectively.
\end{abstract}

\section{Introduction \label{sec:intro}}
Modern AI applications such as retrieval-augmented generation (RAG)~\cite{lewis2020rag}, semantic search~\cite{huang2020fbsearch}, and recommendation~\cite{covington2016youtube} rely on approximate nearest neighbor (ANN) search over embedding vectors~\cite{hedrarag,flowann,rummy}.
These searches often come with a \emph{filter}~\cite{pipeannfilter,adbv,milvus,acorn,curator26,ubase}, such as finding the photos closest to a query photo among those labeled \emph{dog}.
The share of vectors that pass a filter, called its \emph{selectivity}, varies by orders of magnitude between queries~\cite{fannbench,pathseer,fdiskann}.
For large datasets, the vectors and their indexes increasingly reside on SSDs~\cite{diskann,pipeann,smartanns,distvs,spfresh,helmsman}, and the search must still reach a high \emph{recall} (e.g., 0.99)~\cite{pipeann,pipeannfilter,darth,micronn,iks}.

Most on-SSD ANN systems build graph indexes~\cite{diskann,starling,pipeann,odinann}, where a search follows links from vector to vector and reads one SSD record per visited vector.
Unfortunately, under a filter, skipping non-matching vectors can cut the search off from matching vectors.
Although there are repair solutions~\cite{pipeannfilter,fdiskann}, they incur extra SSD reads or index space.

In contrast, a partition index needs no such repair since a query reaches each vector through its partition.
It groups the vectors into partitions, each represented by its centroid, and a query scans only the partitions with the nearest centroids~\cite{faiss}.
However, existing solutions on partition indexes give \emph{the same search budget (i.e., the number of SSD reads) to all queries} in a selectivity range~\cite{adbv} or search route~\cite{milvus,faiss}, or set it by a fixed formula~\cite{pinecone}.
This budget determines the number of partitions to probe and the number of vector candidates to rerank (for correcting errors of compressed codes).
Generally, this value is set to the value of the query requiring the most budget.
Unlike in-memory ANN search, SSD-based ANN search incurs significant overhead because every reranked candidate causes a random I/O read.
As a result, the same budget for all queries can degrade throughput on an SSD by up to an order of magnitude (\S\ref{sec:motiv}).
Existing solutions also apply the filter to far more vectors than a query ranks~\cite{faiss,milvus,adbv,pinecone}.
For example, bitmap filtering marks every match in the dataset even when the query probes a few partitions, adding extra CPU overhead.
In this context, we ask: \emph{can on-SSD vector search achieve high recall for filters of any selectivity while keeping its SSD reads and filtering work low?}

To answer this question affirmatively, we present \sys{}, a filtered vector search system for SSDs that reduces SSD reads while maintaining high recall.
\sys{} sets each query's read budget based on the number of vectors that satisfy its filter, which we call the \emph{match count}.
Our key insight is that, surprisingly, the match count predicts how many SSD reads a query needs for high recall even before it issues a single read.
This is because 1) a query ranks its matches by in-memory compressed codes and reads only the top candidates from the SSD to compute their exact distances, and 2) the number of wrongly ranked matches among these candidates depends on the match count.

A query in \sys{} passes through four steps, and three of them apply this insight.
First, it keeps the matches of each label sorted by where their compressed codes lie in memory (\S\ref{sec:design-filter}).
The length of this list gives the count before the search, and two binary searches give the matches of each probed partition.
Second, it takes the setting for its count from a \emph{plan}, a small table of the fastest settings measured offline to meet the target recall (\S\ref{sec:design-router}).
Third, it stops reading once unread candidates are unlikely to beat the $k$ best read so far, using compression errors recorded per match count (\S\ref{sec:design-d3}).
The remaining step ranks the compressed codes of the matches, prefetching them in the order of the sorted list (\S\ref{sec:design-scan}).

We implement \sys{} and evaluate it against existing graph-based and partition-based solutions on various datasets.
Our experimental results demonstrate the efficiency of \sys.
For example, \sys{} outperforms \pf{}~\cite{pipeannfilter} and \faiss{}~\cite{faiss} by up to 34.5$\times$ and 9.10$\times$, respectively, at equal recall on 100M vectors.

In summary, this work makes the following contributions:
\begin{itemize}[noitemsep,leftmargin=*]
\item We show that, in a partition index, the match count can predict the SSD reads of a query and the error of its compressed codes.
\item We design \sys, which sets the reads of each query from its match count, avoiding unnecessary SSD reads in on-SSD filtered vector search.
\item We show that \sys outperforms graph-based and partition-based systems at high recall.
\end{itemize}

\section{Background and Motivation}
\label{sec:motiv}

\label{sec:motiv-cost}
\begin{figure}[t]
  \centering
  \includegraphics[width=0.95\linewidth]{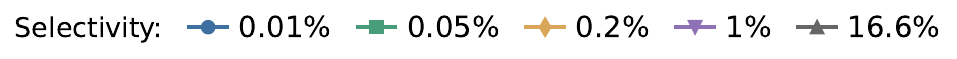}\\[1pt]
  \subfloat[\pfpost{} on SIFT-1B]{\includegraphics[width=0.49\linewidth]{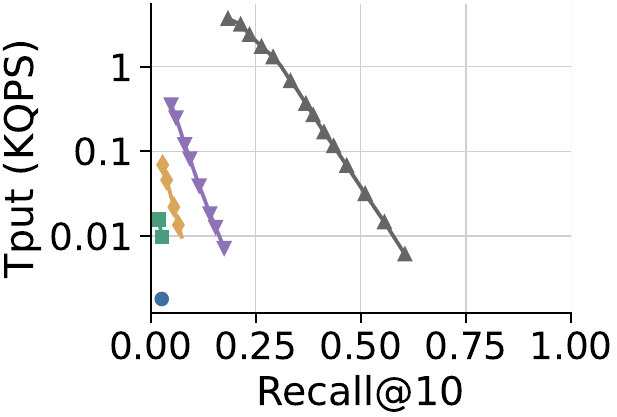}}\hfill
  \subfloat[\fdiskann{} on LAION-100M]{\includegraphics[width=0.48\linewidth]{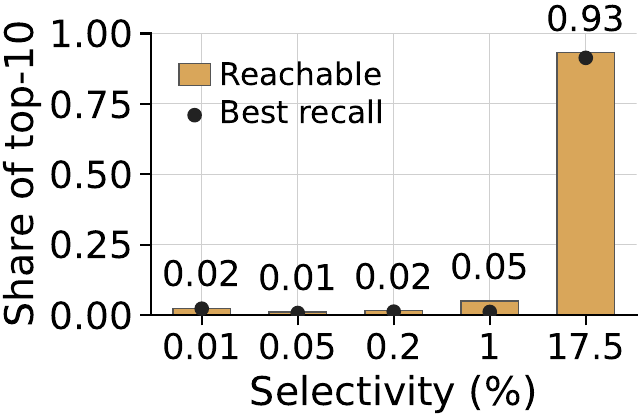}}
  \caption{Filtering a graph on SSDs. (a) Post-filtering fills its candidate list with non-matching vectors, and its recall stays low. (b) For selective labels, few true neighbors are reachable from the entry vectors of their label, and recall stays near zero.}
  \label{fig:graphfail}
\end{figure}

\subsection{Vector Search on SSDs}
\label{sec:motiv-bg}

\emph{Graph indexes} and \emph{partition indexes} are the two dominant index families for approximate nearest neighbor search.
Here, we describe how each index family works and how it processes queries on SSDs.

\noindent\textbf{Graph indexes.}
A graph index links each vector to dozens of nearby vectors~\cite{diskann}.
On an SSD, it stores each raw vector with its links in one SSD record and keeps a short compressed code of every vector in memory~\cite{diskann,starling,pipeann}.
Product quantization (PQ)~\cite{pq} forms this code, called a \emph{PQ code}.
A search starts at an entry vector and hops to vectors ever closer to the query vector.
At each hop, it reads the records of the closest unvisited candidates, ranks their neighbors by PQ codes, and reranks the visited vectors by their raw vectors.
Therefore, each search issues a chain of small dependent reads, as in other traversals on SSDs~\cite{flashgraph,xrp}.

\noindent\textbf{Partition indexes.}
A partition index, or inverted file (IVF) index, groups the vectors into partitions around k-means centroids~\cite{faiss}.
A query compares its vector with the centroids and probes the nearest partitions.
In memory, it either compares the raw vectors in them or ranks their PQ codes.
With PQ codes, it may rerank its \emph{pool} of best candidates with their raw vectors.
Its \emph{budgets}, such as the number of probed partitions and the pool size, set how far it searches and how many candidates it reranks, and larger budgets raise recall at a higher cost.
On an SSD, the PQ codes stay in memory and the raw vectors on the SSD~\cite{zoom,grip,fusionanns}.
A query must thus rerank its pool, and each candidate costs a random read.
At a high target recall, a query reads hundreds of candidates or more~\cite{rabitqssd}.

\subsection{Limitations of Filtered Search on SSDs}

\noindent\textbf{Filtered vector search.}
In production, queries often come with predicates, such as label predicates that require all query labels (AND) or at least one (OR)~\cite{pipeannfilter,fdiskann}.
Filtered vector search returns the $k$ vectors nearest to a query among its \emph{matches}, the vectors that satisfy its predicate~\cite{fannbench,acorn}.
The number of matches is the \emph{match count}, or \emph{count} for short, and their share of all vectors is the \emph{selectivity}.
For example, a query for the label \emph{cat}, which 1\% of 100M vectors carry, has a selectivity of 1\% and a count of 1M.
Our experiments group single-label queries into selectivity \emph{bands}, and this query falls in the 1\% band.
Here, we show why existing vector search on SSDs falls short under such filters.

\noindent\textbf{Graph indexes lose connectivity under a filter.}
A common way to filter a graph search is \emph{post-filtering}, which reads vectors and discards the non-matching ones, wasting their SSD reads.
We use \pfpost{} to denote \pipeann{}~\cite{pipeann}, a recent on-SSD graph system, run with its post-filtering path~\cite{pipeannfilter} forced.
Figure~\ref{fig:graphfail}~(a) shows that its recall stays below 0.9 at every band, while its throughput falls by orders of magnitude.
Skipping non-matching vectors during the traversal would avoid wasted reads, but it cuts the paths of the graph, making many matches unreachable as illustrated in Figure~\ref{fig:insight}~(a).

On-SSD graph systems repair these paths in two ways, and both are costly.
\emph{Speculative filtering} in \pf{}~\cite{pipeannfilter} traverses a superset of the matches with Bloom filters in memory, whose false positives cost SSD reads, and densifies each record with 2-hop neighbors, which we project to grow a billion-scale index over tenfold.
\emph{Label-aware indexing} in \fdiskann~\cite{fdiskann} builds the graph with the labels and an entry vector per label, and its search follows only the neighbors that carry a query label.
Under selective labels, pruning disconnects most matches from the entry vector of their label, and recall stays near zero as shown in Figure~\ref{fig:graphfail}~(b).

\begin{figure}[!t]
  \centering
  \includegraphics[scale=0.52]{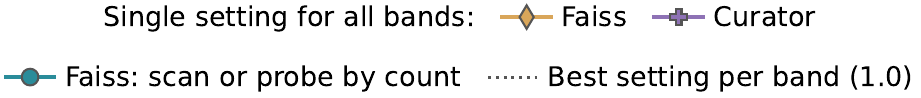}\\[1pt]
  \subfloat[Target recall 0.9]{\includegraphics[width=0.48\linewidth]{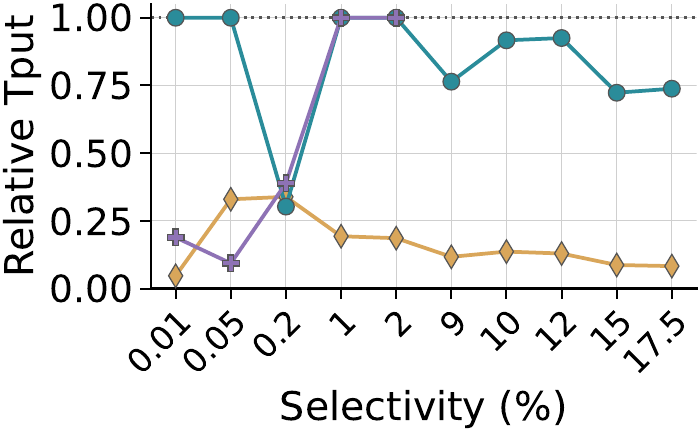}}\hfill
  \subfloat[Target recall 0.99]{\includegraphics[width=0.48\linewidth]{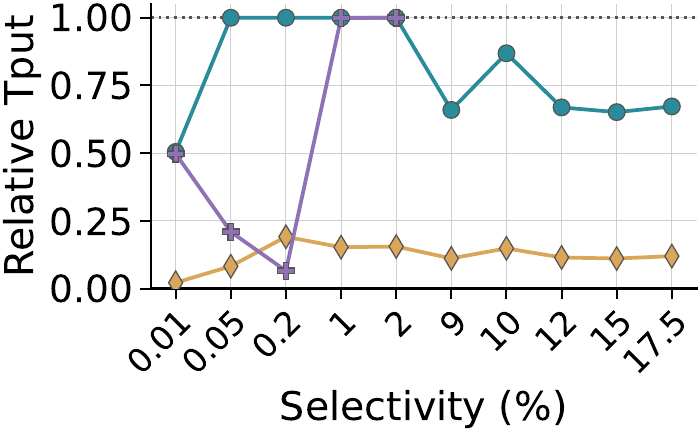}}
  \caption{A single setting for all selectivity bands reaches only a fraction of the throughput of the best setting per band (LAION-100M). We show \curator{} only up to 2\% because it targets low selectivity~\cite{curator26}.}
  \label{fig:naive}
\end{figure}

\begin{table}[t] \centering \caption{Comparison to existing designs.} \label{tab:comparison} \footnotesize \setlength{\tabcolsep}{3pt} \begin{tabular}{@{}lccc@{}} \toprule & \shortstack{Graph\strut\\\cite{pipeann,pipeannfilter,fdiskann}\strut} & \shortstack{Partition\strut\\\cite{faiss, curator26, milvus, adbv}\strut} & \sys{} \\ \midrule Reads per query set by & \shortstack{Graph\strut\\traversal\strut} & \shortstack{Selectivity\strut\\range\strut} & \shortstack{Query's\strut\\match count\strut} \\ \midrule High recall for any filter & $\triangle$ & \cmark & \cmark \\ Low SSD I/O & \xmark & \xmark & \cmark \\ Low filtering cost & \xmark & \xmark & \cmark \\
\bottomrule \end{tabular}
\\[2pt] {\scriptsize $\triangle$: in our runs, \pf{} only by forcing a scan of all matches, \fdiskann{} not for selective labels. \pf{} sizes its pool by its selectivity estimate, and \fdiskann{} and \curator{} (one label) filter cheaply.}
\end{table}

\noindent\textbf{Partition indexes over-provision, at a high cost on SSDs.}
Many filtered search systems build on partition indexes~\cite{faiss, curator26, milvus, adbv}.
Unlike graph indexes, a partition index is not broken by a filter, because a query reaches a match without passing through non-matching vectors.
However, their search budget (i.e., the number of partitions to probe and the number of candidates to rerank) does not follow the match count of each query.
For example, \faiss{}~\cite{faiss} uses the same probe count and rerank pool size for every query.
Extra candidates cost little when raw vectors are in memory, but on an SSD each extra candidate is a random read~\cite{rabitqssd}.

Since partition-based on-SSD ANN systems~\cite{spann,rabitqssd} do not support filtering, we extend \faiss{} and \curator{}~\cite{curator26} to support SSDs as in \S\ref{sec:motiv-bg}.
For \faiss{}, we move the raw vectors used for reranking to the SSD.
For \curator{}, we replace raw-vector distances with PQ estimates during partition search and rerank the final candidates in its fixed-capacity list using raw vectors from the SSD.
Figure~\ref{fig:naive} shows their throughput at each selectivity band.
In some bands it falls below a tenth of the throughput with the best setting because a single setting for all bands must allow enough reads for the hardest query.
As the Faiss authors suggest, a query with few matches can scan all matches (the \emph{exhaustive} route) instead of probing partitions (the \emph{partition} route), and choosing this route by the count recovers much of the loss.
Even so, all queries within a route use the same rerank pool sized for the query that needs the most reads, so easier queries still waste SSD reads.
Similarly, budgets set by the selectivity follow the count more closely, but they still over-provision for the query in their range that needs the most reads~\cite{adbv} or follow a fixed formula~\cite{pinecone}.

At the same time, these systems incur CPU overhead by filtering more vectors than a query needs, further reducing throughput.
Bitmap filtering marks every match of a query even when the query probes a few partitions~\cite{milvus,adbv,pinecone}, and inline filtering tests every probed vector~\cite{faiss}.
Lists per label~\cite{ivf2,curator26} avoid this cost for a single label, but for an AND predicate they first compute all its matches (e.g., by intersecting the label lists).
Table~\ref{tab:comparison} compares these designs.

\subsection{Our Approach}
\label{sec:motiv-approach}

Our high-level idea is to set the SSD reads of each query from its match count on a partition index.

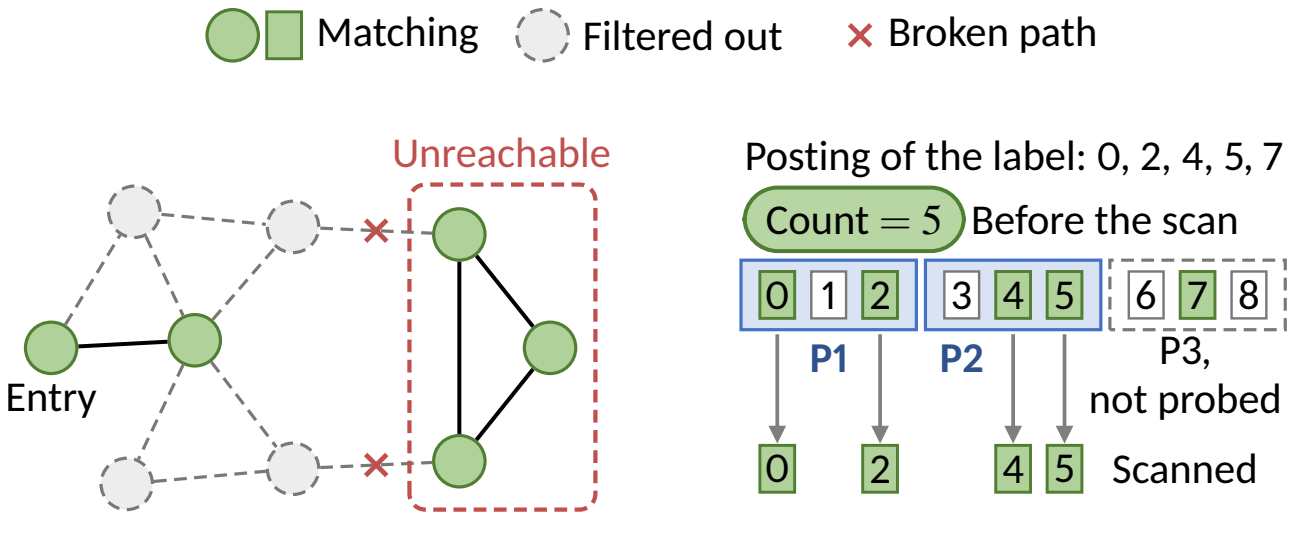
\begin{figure}[t]
\centering
  \begin{tikzpicture}[kfig]
\node[kgm] at (1.0,3.62) {};
\node[kcodeh] at (1.34,3.62) {};
\node[ktxt, anchor=west] at (1.52,3.62) {Matching};
\node[kgx] at (3.05,3.62) {};                 \node[ktxt, anchor=west] at (3.28,3.62) {Filtered out};
\draw[kRed, line width=0.9pt] (5.08,3.55) -- ++(0.14,0.14) (5.08,3.69) -- ++(0.14,-0.14);
\node[ktxt, anchor=west] at (5.3,3.62) {Broken path};
\end{tikzpicture}\\[2pt]
  \subfloat[Graph index]{\begin{tikzpicture}[kfig]
\useasboundingbox (0.1,0.62) rectangle (3.98,3.18);
\node[kgm] (e) at (0.35,1.75) {};
\node[ktxt, anchor=north] at (0.35,1.55) {Entry};
\node[kgm] (a) at (1.3,1.8) {};
\node[kgx] (g1) at (0.9,2.65) {};
\node[kgx] (g2) at (1.95,2.55) {};
\node[kgx] (g3) at (0.85,0.85) {};
\node[kgx] (g4) at (1.95,0.95) {};
\node[kgm] (m1) at (3.05,2.5) {};
\node[kgm] (m2) at (3.65,1.75) {};
\node[kgm] (m3) at (3.05,1.0) {};
\draw[kgd] (e) -- (g1) (g1) -- (g2) (g1) -- (a) (a) -- (g2) (a) -- (g3) (g3) -- (g4) (a) -- (g4);
\draw[kgd] (g2) -- (m1) (g4) -- (m3);
\draw[kge] (e) -- (a) (m1) -- (m2) (m2) -- (m3) (m3) -- (m1);
\foreach \u/\v in {g2/m1, g4/m3}
  \draw[kRed, line width=0.9pt] ($(\u)!0.5!(\v)+(-0.07,-0.07)$) -- ++(0.14,0.14) ($(\u)!0.5!(\v)+(-0.07,0.07)$) -- ++(0.14,-0.14);
\draw[kRed, line width=0.8pt, kdash, rounded corners=3pt] (2.72,0.68) rectangle (3.95,2.83);
\node[ktxt, text=kRed, anchor=south] at (3.33,2.91) {Unreachable};
\end{tikzpicture}}\hfill
  \subfloat[Partition index]{\begin{tikzpicture}[kfig]
\useasboundingbox (4.30,0.62) rectangle (7.95,3.18);
\def\cw{0.34}
\def\xo{4.45}
\foreach \pp/\on in {0/1, 1/1, 2/0} {
  \pgfmathsetmacro{\xa}{\xo-0.07+\pp*(3*\cw+0.2)}
  \ifnum\on=1 \draw[kpartn] (\xa,1.87) rectangle ++(3*\cw+0.14,0.46);
  \else \draw[kpartx] (\xa,1.87) rectangle ++(3*\cw+0.14,0.46); \fi
}
\foreach \i in {0,...,8} {
  \pgfmathsetmacro{\xx}{\xo+\cw/2+\i*\cw+floor(\i/3)*0.2}
  \def\st{kcodep}
  \foreach \m in {0,2,4,5,7} {\ifnum\i=\m \global\def\st{kcodeh}\fi}
  \node[\st] (c\i) at (\xx,2.1) {\i};
  \global\def\st{kcodep}
}
\node[ktxt, anchor=west] at (\xo-0.08,3.0) {Posting of the label: 0, 2, 4, 5, 7};
\node[kpart={kGreenS}{kGreen}, kpill, minimum height=4.2mm, inner xsep=4pt, anchor=west] (cnt) at (\xo-0.06,2.6) {Count $=5$};
\node[ktxt, anchor=west] at (cnt.east) {Before the scan};
\node[khdr, text=kBlueS] at (c1 |- 0,1.68) {P1};
\node[khdr, text=kBlueS] at (c3 |- 0,1.68) {P2};
\node[ktxt] at (7.32,1.55) {P3,\\not probed};
\foreach \i in {0,2,4,5} {
  \node[kcodeh] (s\i) at (c\i |- 0,0.95) {\i};
  \draw[kar] (c\i |- 0,1.84) -- (s\i.north);
}
\node[ktxt, anchor=west] at (6.8,0.95) {Scanned};
\end{tikzpicture}}
\caption{Example of the impact of filtering. (a) Removing graph nodes breaks the paths to matching vectors. (b) A partition probe ranks only the matching PQ codes of its partitions.}
  \label{fig:insight}
\end{figure}

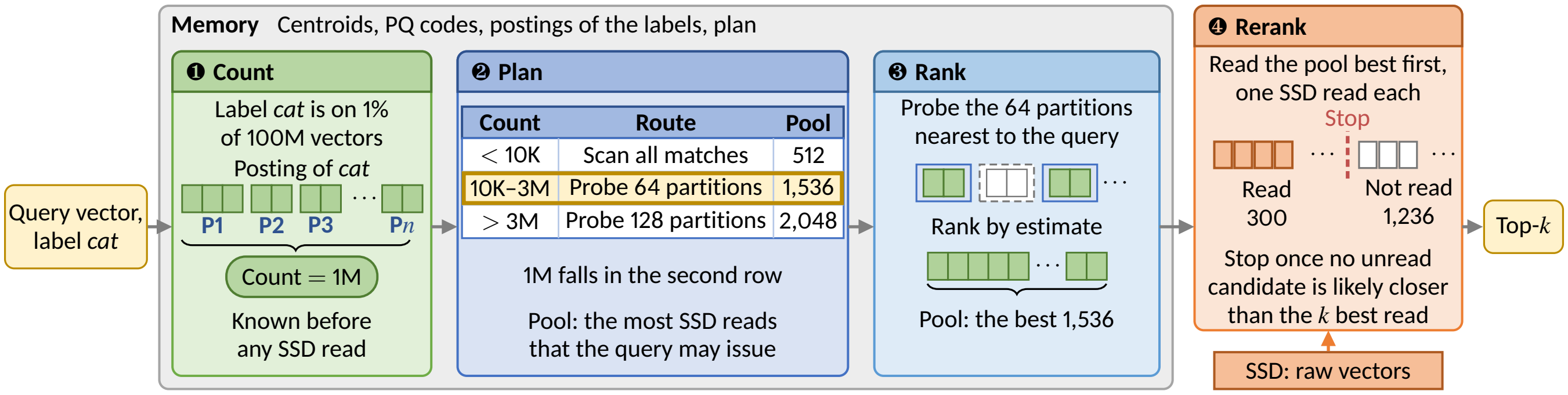
\begin{figure*}[t]
  \centering
  \begin{tikzpicture}[kfig]
\def\cw{0.23}
\draw[kregion={kGray}{kGrayL}, kround] (1.7,0.9) rectangle (13.25,5.27);
\node[ktxt, anchor=west] at (1.8,5.04) {\textbf{Memory}\hspace{6pt}Centroids, PQ codes, postings of the labels, plan};
\node[kcard={kGoldS}{kGoldL}, kround, minimum width=1.5cm, minimum height=0.95cm] (q) at (0.75,2.75) {Query vector,\\label \emph{cat}};
\node[kcard={kGoldS}{kGoldL}, kround, minimum width=0.9cm, minimum height=0.6cm] (topk) at (17.25,2.75) {Top-$k$};
\kstage{1.85}{1.05}{2.95}{3.7}{kGreen}{kGreenS}{kGreenL}{1}{Count}
\kstage{5.1}{1.05}{4.45}{3.7}{kBlue}{kBlueS}{kBlueL}{2}{Plan}
\kstage{9.85}{1.05}{3.25}{3.7}{kSky}{kSkyS}{kSkyL}{3}{Rank}
\kstage{13.5}{1.57}{3.05}{3.7}{kOrange}{kOrangeS}{kOrangeL}{4}{Rerank}
\node[ktxt] at (3.325,3.93) {Label \emph{cat} is on 1\%\\of 100M vectors};
\node[ktxt] at (3.325,3.4) {Posting of \emph{cat}};
\begin{scope}[shift={(2.075,3.07)}]
  \foreach \x in {0,1,2} \node[kcodeh] at (\x*\cw,0) {};
  \foreach \x in {0,1} \node[kcodeh] at (0.79+\x*\cw,0) {};
  \foreach \x in {0,1} \node[kcodeh] at (1.35+\x*\cw,0) {};
  \node[ktxt] at (1.98,0) {$\cdots$};
  \foreach \x in {0,1} \node[kcodeh] at (2.28+\x*\cw,0) {};
  \foreach \c/\l in {0.23/P1, 0.905/P2, 1.465/P3, 2.395/P$n$} \node[khdr, text=kBlueS] at (\c,-0.3) {\l};
  \draw[kbrace] (2.48,-0.47) -- (-0.12,-0.47);
\end{scope}
\node[kpart={kGreenS}{kGreen}, kpill, minimum height=4.6mm, inner xsep=5pt] (cnt) at (3.325,2.18) {Count $=$ 1M};
\node[ktxt] at (3.325,1.5) {Known before\\any SSD read};
\begin{scope}[shift={(5.15,2.62)}]
  \draw[kpart={kBlueS}{kBlue}] (0,1.14) rectangle (4.35,1.5);
  \node[khdr] at (0.55,1.32) {Count};
  \node[khdr] at (2.325,1.32) {Route};
  \node[khdr] at (3.95,1.32) {Pool};
  \draw[kcard={kBlue}{white}] (0,0) rectangle (4.35,1.14);
  \fill[kGoldL] (0,0.38) rectangle (4.35,0.76);
  \foreach \y in {0.76,0.38} \draw[kBlueM, line width=0.5pt] (0,\y) -- (4.35,\y);
  \foreach \x in {1.1,3.55} \draw[kBlueM, line width=0.5pt] (\x,0) -- (\x,1.14);
  \foreach \y/\c/\r/\p in {0.95/{$<$ 10K}/{Scan all matches}/{512}, 0.57/{10K--3M}/{Probe 64 partitions}/{1,536},
                           0.19/{$>$ 3M}/{Probe 128 partitions}/{2,048}} {
    \node[ktxt] at (0.55,\y) {\c}; \node[ktxt] at (2.325,\y) {\r}; \node[ktxt] at (3.95,\y) {\p};
  }
  \draw[ksel] (0.02,0.4) rectangle (4.33,0.74);
\end{scope}
\node[ktxt] at (7.325,2.2) {1M falls in the second row};
\node[ktxt] at (7.325,1.5) {Pool: the most SSD reads\\that the query may issue};
\node[ktxt] at (11.475,3.93) {Probe the 64 partitions\\nearest to the query};
\begin{scope}[shift={(10.33,3.25)}]
  \draw[kpartn] (0,-0.21) rectangle (0.62,0.21);
  \foreach \x in {0,1} \node[kcodeh] at (0.195+\x*\cw,0) {};
  \draw[kpartx] (0.72,-0.21) rectangle (1.34,0.21);
  \foreach \x in {0,1} \node[kcodep] at (0.915+\x*\cw,0) {};
  \draw[kpartn] (1.44,-0.21) rectangle (2.06,0.21);
  \foreach \x in {0,1} \node[kcodeh] at (1.635+\x*\cw,0) {};
  \node[ktxt] at (2.28,0) {$\cdots$};
\end{scope}
\node[ktxt] at (11.475,2.74) {Rank by estimate};
\begin{scope}[shift={(10.575,2.3)}]
  \foreach \x in {0,...,4} \node[kcodeh] at (\x*\cw,0) {};
  \node[ktxt] at (1.27,0) {$\cdots$};
  \foreach \x in {0,1} \node[kcodeh] at (1.58+\x*\cw,0) {};
  \draw[kbrace] (1.92,-0.2) -- (-0.12,-0.2);
\end{scope}
\node[ktxt] at (11.475,1.68) {Pool: the best 1,536};
\node[ktxt] at (15.025,4.45) {Read the pool best first,\\one SSD read each};
\begin{scope}[shift={(13.825,3.57)}]
  \foreach \x in {0,...,3} \draw[kpart={kOrangeS}{kOrange}] (\x*\cw-0.1,-0.15) rectangle ++(0.2,0.3);
  \node[ktxt] at (1.15,0) {$\cdots$};
  \draw[kRed, line width=1.2pt, kdash] (1.42,-0.27) -- (1.42,0.27);
  \foreach \x in {0,1,2} \draw[kcard={kGrayS}{white}] (1.65+\x*\cw-0.1,-0.15) rectangle ++(0.2,0.3);
  \node[ktxt] at (2.53,0) {$\cdots$};
  \node[ktxt, text=kRed] at (1.42,0.4) {Stop};
  \node[ktxt] at (0.5,-0.55) {Read\\300};
  \node[ktxt] at (2.12,-0.55) {Not read\\1,236};
\end{scope}
\node[ktxt] at (15.025,2.07) {Stop once no unread\\candidate is likely closer\\than the $k$ best read};
\node[kpart={kOrangeS}{kOrange}, minimum width=2.6cm, minimum height=0.42cm] (ssd) at (15.025,1.11) {SSD: raw vectors};
\draw[karrow=kOrange] (ssd.north) -- (15.025,1.57);
\draw[karr] (q.east) -- (1.85,2.75);
\draw[karr] (4.8,2.75) -- (5.1,2.75);
\draw[karr] (9.55,2.75) -- (9.85,2.75);
\draw[karr] (13.1,2.75) -- (13.5,2.75);
\draw[karr] (16.55,2.75) -- (topk.west);
\end{tikzpicture}
  \caption{\sys{} pipeline with an example query for the label \emph{cat}, which 1\% of 100M vectors carry. Only the rerank reads the SSD, and it stops long before the end of the pool.}
  \label{fig:arch}
\end{figure*}

\noindent\textbf{Why the count sets the reads.}
The count tells how many candidates a query must read.
A query ranks its candidates by the PQ codes, whose compression errors can make some candidates appear closer to the query than the true neighbors.
When more vectors match, there are likely to be more such candidates, so the query tends to read more candidates to reach the same recall.
This explains why the best setting in Figure~\ref{fig:naive} changes with the band.
However, the number of reads for a given count is hard to predict with a formula.
This is because the recall of a fixed setting changes with the count, in opposite directions on the partition and exhaustive routes (\S\ref{sec:design-plan}).

\noindent\textbf{Finding the count and the matches.}
A partition index can give this count before any SSD read.
Figure~\ref{fig:insight}~(b) shows an example with nine vectors in three partitions.
Their PQ codes lie in memory in partition order, numbered 0 to 8.
For each label, we keep a sorted list of the numbers of the vectors that carry it, called the label's \emph{posting}.
In the example, the posting of the label is 0, 2, 4, 5, 7.
Its length gives the count of five before any SSD read.
In addition, the posting shows the matches in each partition, because each partition holds consecutive numbers.
Therefore, a probe of P1 and P2 ranks only the PQ codes 0, 2, 4, and 5 and checks no other vector.

\noindent\textbf{Measured settings and calibrated stops.}
We measure the fastest setting that meets the target at a few counts and keep these settings in a small table.
In addition, we also measure the errors of the PQ codes at each count, and a query stops reading once the unread candidates are unlikely to hold a true neighbor.
As a result, each query reads far fewer vectors than under a single setting.

\section{\sys{} Design}
\label{sec:design}

\subsection{Overview}
\label{sec:design-layout}

\noindent\textbf{Mechanisms.}
\sys{} serves each query in four steps, which Figure~\ref{fig:arch} shows on one query.
It first obtains the count and the matches of the query from its postings, and then takes a setting for this count from the plan.
Next, it ranks the PQ codes of the matches in memory, and lastly it reranks the best candidates with SSD reads until it can stop.
The PQ codes of all vectors lie in one array in partition order, and the SSD holds the raw vectors.
\begin{itemize}[noitemsep,leftmargin=*]
\item \textbf{Position-ordered postings (\S\ref{sec:design-filter}).}
We call the index of a PQ code in the array its \emph{position}.
The posting of each label lists the sorted positions of the vectors that carry it.
Its length gives the count, and two binary searches give the matches in each probed partition.
Attribute values and vector identifiers sit in arrays indexed by position.
\item \textbf{Count-indexed plan (\S\ref{sec:design-router}).}
The plan is a small table whose rows cover ranges of the count.
Each row fixes the route, the budgets, and when to stop reading, and calibration measures the recall of every row offline.
\item \textbf{Lookahead scan (\S\ref{sec:design-scan}).}
The scan ranks the matching PQ codes and fetches those of later candidates while it computes the current one.
\item \textbf{Calibrated stopping (\S\ref{sec:design-d3}).}
The rerank reads the best candidates from the SSD, one aligned read each, and stops with error bounds measured at the count of the query.
\end{itemize}

\noindent\textbf{Example.}
Figure~\ref{fig:arch} follows one query for the label \emph{cat}, which 1\% of 100M vectors carry.
The length of the posting of \emph{cat} gives a count of 1M (\ding{182}).
This count falls in the second row of the plan, which probes the 64 nearest partitions with a pool of 1,536 candidates (\ding{183}).
The probe ranks the PQ codes of the matches in these partitions in memory and keeps the best 1,536 (\ding{184}).
Lastly, the rerank reads these candidates from the SSD in rank order and stops once no unread candidate is likely closer than the $k$ best read so far (\ding{185}).
In this example, it stops after 300 reads, and \sys{} returns the $k$ nearest vectors that it read.

\subsection{Predicates on Position-Ordered Postings}
\label{sec:design-filter}

In the first step, a query obtains its count and the matches in each partition that it probes, before it computes any distance.
The posting of each label gives the count by its length, but a posting that simply lists vector identifiers does not tell which matches lie in a probed partition.
A probe must then mark all matches in a bitmap or check the labels of every PQ code that it visits, which wastes work even when the query probes only a few partitions.
\sys{} instead keeps the posting of each label sorted by position (i.e., by where the PQ codes of its matches lie in memory).
Because the PQ codes of one partition lie next to each other, the matches of a partition form one contiguous run of the posting.
Two binary searches for the boundaries of a partition therefore give the matches of each probed partition without a per-query bitmap.
In addition, the exhaustive route reads the same postings in position order, so it accesses the PQ codes forward through memory rather than at random.

\sys{} evaluates a multi-label predicate inside each probed partition rather than over the whole dataset.
It intersects the runs of an AND starting from the rarest label, and it merges the runs of an OR and removes duplicates.
Because the result stays in position order, the filtering work follows only the partitions that the query probes.

\noindent\textbf{Counting the matches.}
For a single label, the count is the length of its posting.
For an OR, it is the sum of the posting lengths, an upper bound that calibration also uses.
For an AND, an exact count must intersect the postings, which takes time that grows with the length of the rarest posting.
However, \sys{} uses the count only to select a row of its plan, a table whose rows each cover a range of counts, so the count need not be exact.
Therefore, when the rarest posting is long in an AND, \sys{} instead estimates the count with a confidence interval.
It samples this posting at even spacing and counts how many samples also carry the other labels.
For example, if 10\% of 16K samples from a posting of 1M entries match, the interval places the count between 93K and 107K.
\sys{} intersects the postings exactly only when this interval spans more than one row.

\noindent\textbf{Other predicates.}
The same per-partition evaluation handles ranges on numeric ranges, string conditions, negation, and nesting.
Inside a probed partition, \sys{} takes candidates from the operand (e.g., a label or a range) with the fewest of them.
The candidates thus come from a label's posting if the predicate has a label, and from every position of the partition otherwise.
\sys{} then tests each candidate against the whole predicate in position order.
Over the whole dataset (e.g., on the exhaustive route), a selective range or string condition takes its candidates from a copy of its column sorted by value.
\sys{} bounds the count of such a predicate from the counts of its parts.
It counts exactly, first by sampling as above, only when the bounds select different plan rows.

\subsection{Count-Indexed Planning}
\label{sec:design-router}
\label{sec:design-plan}

In the second step, the query takes the cheapest route and budgets that still meet its target recall at its count.
A cost model could choose a setting for each count, but no formula that we tested predicts the recall of a filtered setting on an SSD (\S\ref{sec:eval-deep}).
\sys{} therefore reads them from the plan, a small table of the fastest settings that meet the target, measured offline at a few counts.
These few counts suffice, since a setting that meets the target at two counts also serves the counts between them.
This is because the recall of a fixed setting on each route moves in one direction as the count changes, for labels whose matches spread alike.
Recall falls on the exhaustive route as the count grows and on the partition route as the count shrinks.

\noindent\textbf{The plan.} %
Each row of the plan covers a range of counts and fixes the route, the number of probed partitions, and the pool size.
In addition, a row that stops early fixes the stopping table of \S\ref{sec:design-d3} and its quantile level.
At query time, \sys{} selects the row whose range contains the count, which costs a few comparisons.
The plan works like a parametric plan in a database optimizer, but is indexed by the count that the postings give before the first read.
Unlike a parametric plan, every row must meet a target recall rather than only minimize a cost, and \sys{} measures the recall of every row rather than estimating it.

\begin{figure}[t]
\centering
  \begin{tikzpicture}[kfig]
\def\w{0.95}
\draw[karr] (0.05,0.62) -- (7.55,0.62);
\node[ktxt, anchor=south east] at (7.5,0.66) {Count};
\foreach \c/\x/\col in {1/1.5/kSkyS, 2/3.15/kSkyS, 3/4.8/kBlueS, 4/6.45/kBlueS} {
  \foreach \f in {0,0.5,1} {
    \draw[\col, line width=0.45pt, kdash] (\x-\w+\f*\w,0.62) -- (\x-\w+\f*\w,1.08);
    \fill[\col] (\x-\w+\f*\w,0.62) circle[radius=0.55mm];
  }
  \node[ktxt, anchor=north] at (\x-\w/2,0.52) {Point \c};
}
\draw[kregion={kSkyS}{kSkyL}] (0.15,1.08) rectangle (1.5,2.0);
\draw[kpart={kSkyS}{kSky}] (1.5-\w,1.08) rectangle (1.5,2.0);
\draw[kpart={kSkyS}{kSky}] (3.15-\w,1.08) rectangle (3.15,2.0);
\draw[kpart={kBlueS}{kBlue}] (4.8-\w,1.08) rectangle (4.8,2.0);
\draw[kregion={kBlueS}{kBlueL}] (6.45-\w,1.08) rectangle (7.25,2.0);
\draw[kpart={kBlueS}{kBlue}] (6.45-\w,1.08) rectangle (6.45,2.0);
\foreach \x in {1.5,3.15,4.8,6.45} \node[ktxt] at (\x-\w/2,1.54) {Stops\\early};
\draw[kar=kSkyS] (0.5,1.54) -- (0.2,1.54);
\draw[kar=kBlueS] (6.55,1.54) -- (7.2,1.54);
\foreach \a/\b in {1.5/2.2, 4.8/5.5} {
  \draw[kregion={kGrayS}{kGrayL}] (\a,1.08) rectangle (\b,2.0);
  \node[ktxt] at ({(\a+\b)/2},1.54) {Fixed};
}
\draw[kregion={kGrayS}{kGoldL}] (3.15,1.08) rectangle (3.85,2.0);
\draw[ksel] (3.17,1.1) rectangle (3.83,1.98);
\node[ktxt] at (3.5,1.54) {Fixed};
\fill[kGoldS] (3.5,0.62) circle[radius=0.75mm];
\draw[kGoldS, line width=0.8pt] (3.5,0.62) -- (3.5,1.08);
\node[ktxt, anchor=south] at (3.5,2.24) {The row of the query's count};
\draw[kar=kGoldS, line width=0.9pt] (3.5,2.26) -- (3.5,2.02);
\draw[kbrace, decoration={mirror}] (3.15-\w,0.12) -- (4.8,0.12);
\node[ktxt, anchor=north] at (3.5,-0.02) {Fixed row meets the target at all six counts};
\begin{scope}[yshift=-0.62cm]
  \draw[kpart={kSkyS}{kSky}] (0.3,-0.1) rectangle ++(0.3,0.2); \node[ktxt, anchor=west] at (0.66,0) {Exhaustive row of a point};
  \draw[kpart={kBlueS}{kBlue}] (4.25,-0.1) rectangle ++(0.3,0.2); \node[ktxt, anchor=west] at (4.61,0) {Partition row of a point};
\end{scope}
\begin{scope}[yshift=-1.0cm]
  \draw[kregion={kGrayS}{kGrayL}] (0.3,-0.1) rectangle ++(0.3,0.2); \node[ktxt, anchor=west] at (0.66,0) {Gap row, reads its whole pool};
  \fill[kSkyS] (4.25,0) circle[radius=0.55mm]; \fill[kBlueS] (4.42,0) circle[radius=0.55mm];
  \node[ktxt, anchor=west] at (4.53,0) {80, 90, 100\% of a point's count};
\end{scope}
\end{tikzpicture}
\caption{Example of a count-indexed plan. Calibration measures one row at each of four counts (Points 1 to 4) and one fixed row for each gap between them. A query takes the row that covers its count.}
\label{fig:router}
\end{figure}
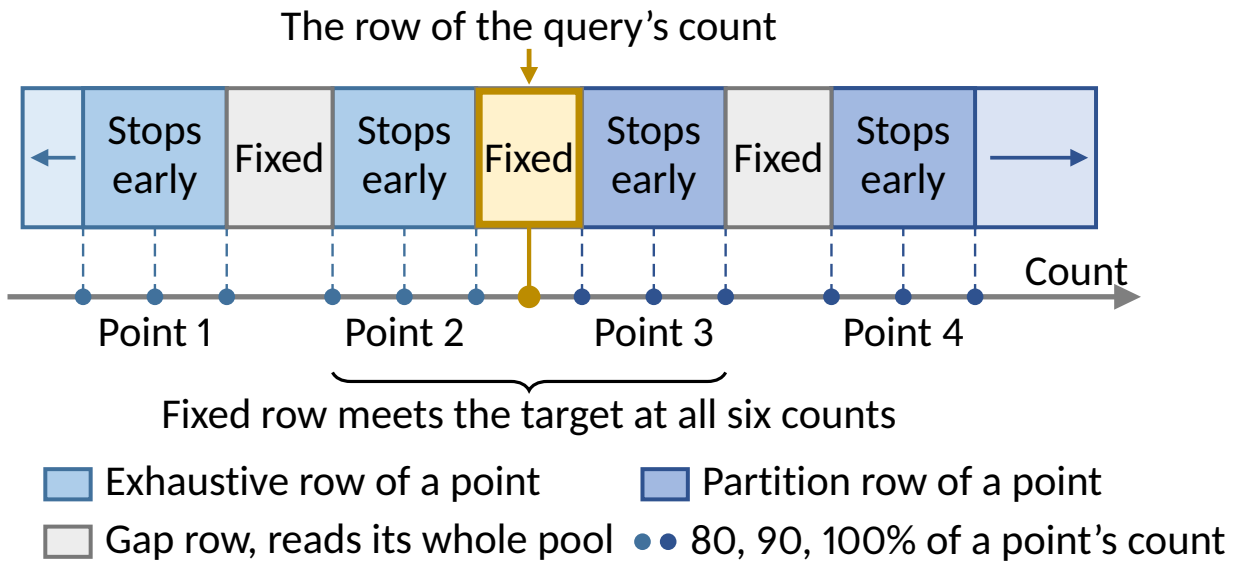

\sys{} builds the plan offline from held-out queries at a few counts, which we call \emph{calibration points}.
For each point and target, it measures candidate settings on both routes and selects the fastest one that meets the target with a margin (Figure~\ref{fig:router}).
The margin subtracts two standard errors from the measured recall, counting each neighbor of a calibration query as one trial.
The margin protects the plan against the sampling error of a finite set of calibration queries.
For example, with 500 calibration queries of 10 neighbors each, a row for the 0.99 target must measure a recall of at least 0.993.

\noindent\textbf{Covering a range of counts.}
When the calibration queries of a point share one label, they cover only one count.
\sys{} therefore also measures each setting on copies of the label's posting that keep a random 90\% and 80\% of its members, so the row of the point covers the range of these three counts.
In these copies, the true neighbors are recomputed over the kept members, since the removed vectors no longer match.
Because these copies remove random members, as random deletions do, the row also keeps serving a label whose count drops after deletions.

Between two calibration points, the row of either point can miss the target near the other point.
This is because recall moves in opposite directions on the two routes.
For such a gap, \sys{} uses the fastest fixed setting that meets the target at the six calibration counts of its two points, without stopping early.
Therefore, this setting holds inside the gap.

\noindent\textbf{Example.}
Figure~\ref{fig:router} shows a plan with four calibration points.
At each point, calibration measures the settings at 80\%, 90\%, and 100\% of its count, drawn as dots on the count axis.
Points 1 and 2 have few matches, and their rows scan all matches.
Points 3 and 4 have many matches, and their rows probe partitions.
The row of each point stops early with the stopping table of that point.
Between Points 2 and 3, a query takes the fixed row of the gap (brace), which reads its whole pool.
Likewise, a query below Point 1 takes the row of Point 1, and a query above Point 4 takes the row of Point 4.

\noindent\textbf{A local count.}
The count says how many vectors match, but not how closely they lie to the query in the vector space.
For example, when the query is a photo of a cat and its filter is the label cat, many vectors labeled cat lie close to the query.
Its matches are then denser around it than around the calibration queries of its count, and the stopping table of that count can end its reads too early.
Therefore, \sys{} can also count the matches in the partitions nearest to the query.
If the label is much denser there than overall, \sys{} uses the row of the count that the label would have at this density everywhere, without recalibration.
For example, a label that holds 1\% of all vectors but 5\% of those in the nearest partitions takes the row of a five times larger count.

\noindent\textbf{Finding the nearest centroids.}
\label{sec:design-coarse}
Because scanning all centroids dominates the partition route when there are many centroids of high dimension, \sys{} searches a graph index over the centroids, which no filter breaks.
For a few thousand centroids, it scans them exactly.
After the build, it links every centroid that the entry point cannot reach to a nearby reached one, because its partition would otherwise never be probed.

\subsection{Scanning with Lookahead Prefetching}
\label{sec:design-scan}

In the third step, the query ranks the PQ codes of its matches in memory.
With postings, the filter touches only the matches, so this ranking becomes the main work in memory and limits the throughput once the reads shrink.
\sys{} therefore prefetches the memory of later candidates while it computes the current one, and backs the arrays of the scan with huge pages.
This works because the posting lists every candidate position in advance, so the scan knows its future accesses.
Without this lookahead, a plain loop waits on memory at almost every candidate, because the matching positions are sorted but sparse and consecutive candidates rarely share a cache line.
Their PQ codes and identifiers also lie in arrays of gigabytes, beyond what the TLB covers with regular pages.

\begin{figure}[t]
  \centering
  \subfloat[Accesses of the scan]{\begin{tikzpicture}[kfig]
\useasboundingbox (-0.05,2.86) rectangle (7.95,5.3);
\node[ktxt, anchor=west, align=left] at (0.0,4.47) {Matching\\positions};
\foreach \p [count=\i from 0] in {3,17,42,58,91,130,162,205} {
  \def\st{kcell}
  \ifnum\i=2 \def\st{khit}\fi
  \ifnum\i=6 \def\st{kpos}\fi
  \node[\st, minimum width=5mm, minimum height=3.2mm, font=\figfontS] (p\i) at (1.75+\i*0.8,4.47) {\p};
}
\node[ktxt] at (7.9,4.47) {\dots};
\draw[draw=black, line width=0.6pt, {Triangle[length=1.2mm,width=1.1mm]}-{Triangle[length=1.2mm,width=1.1mm]}] ($(p2.north)+(0,0.5)$) -- ($(p6.north)+(0,0.5)$);
\node[ktxt, fill=white] at ($(p4.north)+(0,0.5)$) {4 candidates ahead};
\def\xi{2.725}   %
\def\xd{6.225}   %
\draw[kuse] (p2.south) -- (\xi,3.71);
\draw[kpre] (p6.south) -- (\xd,3.71);
\draw[kGreenS, line width=0.6pt, densely dotted] (\xi,3.41) -- (\xi,3.19);
\draw[kSkyS, line width=0.6pt, densely dotted] (\xd,3.41) -- (\xd,3.19);
\foreach \y/\lab in {3.55/PQ codes, 3.05/IDs} {
  \node[ktxt, anchor=west] at (0.0,\y) {\lab};
  \draw[draw=kGrayS, fill=kGrayL, line width=0.5pt] (1.35,\y-0.14) rectangle (7.85,\y+0.14);
  \foreach \t in {1.6,1.85,...,7.7} \draw[kGrayM, line width=0.3pt] (\t,\y-0.14) -- (\t,\y+0.14);
  \draw[khit, minimum width=0pt] (\xi-0.12,\y-0.14) rectangle (\xi+0.13,\y+0.14);
  \draw[draw=kSkyS, fill=kSkyM, line width=0.55pt] (\xd-0.12,\y-0.14) rectangle (\xd+0.13,\y+0.14);
}
\node[ktxt, anchor=west] at (3.15,4.02) {Compute};
\node[ktxt, anchor=west] at (6.6,4.02) {Prefetch};
\end{tikzpicture}}\\[2pt]
  \begin{tikzpicture}[kfig]
\draw[kwait] (2.75,1.38) rectangle ++(0.32,0.2);
\node[ktxt, anchor=west] at (3.12,1.48) {Code and ID wait};
\draw[kcomp] (5.3,1.38) rectangle ++(0.32,0.2);
\node[ktxt, anchor=west] at (5.67,1.48) {Distance compute};
\end{tikzpicture}\\[1pt]
  \subfloat[Without lookahead]{\begin{tikzpicture}[kfig]
\useasboundingbox (-0.05,1.58) rectangle (7.95,2.28);
\foreach \k/\lab in {0/42, 1/58, 2/91} {
  \draw[kwait] (0.3+\k*2.2,1.93) rectangle ++(1.6,0.3);
  \node[ktxt] at (1.1+\k*2.2,2.08) {\lab};
  \draw[kcomp] (1.9+\k*2.2,1.93) rectangle ++(0.6,0.3);
}
\draw[kRed, line width=0.9pt, {Triangle[length=1.2mm,width=1.1mm]}-{Triangle[length=1.2mm,width=1.1mm]}] (2.5,1.8) -- (4.1,1.8);
\node[ktxt, anchor=west] at (4.25,1.73) {Compute idle};
\end{tikzpicture}}\\[2pt]
  \subfloat[With lookahead (\sys{})]{\begin{tikzpicture}[kfig]
\useasboundingbox (-0.05,-0.68) rectangle (7.95,1.3);
\foreach \k/\lab in {0/42, 1/58, 2/91, 3/130} {
  \draw[kwait] (0.3+\k*0.6,0.95-\k*0.36) rectangle ++(1.6,0.3);
  \node[ktxt] at (1.1+\k*0.6,1.1-\k*0.36) {\lab};
  \draw[kcomp] (1.9+\k*0.6,0.95-\k*0.36) rectangle ++(0.6,0.3);
}
\node[ktxt, anchor=west] at (4.85,0.0) {Compute stays busy};
\draw[karr] (0.0,-0.38) -- (7.9,-0.38);
\node[ktxt, anchor=north east] at (7.9,-0.42) {Time};
\end{tikzpicture}}
\caption{Example of lookahead prefetching. (a) While the scan computes the candidate at position 42, it already fetches the PQ code and identifier of position 162. (b) Without lookahead, the CPU waits on memory at every candidate. (c) With lookahead, these waits overlap the computation.}
  \label{fig:lookahead}
\end{figure}
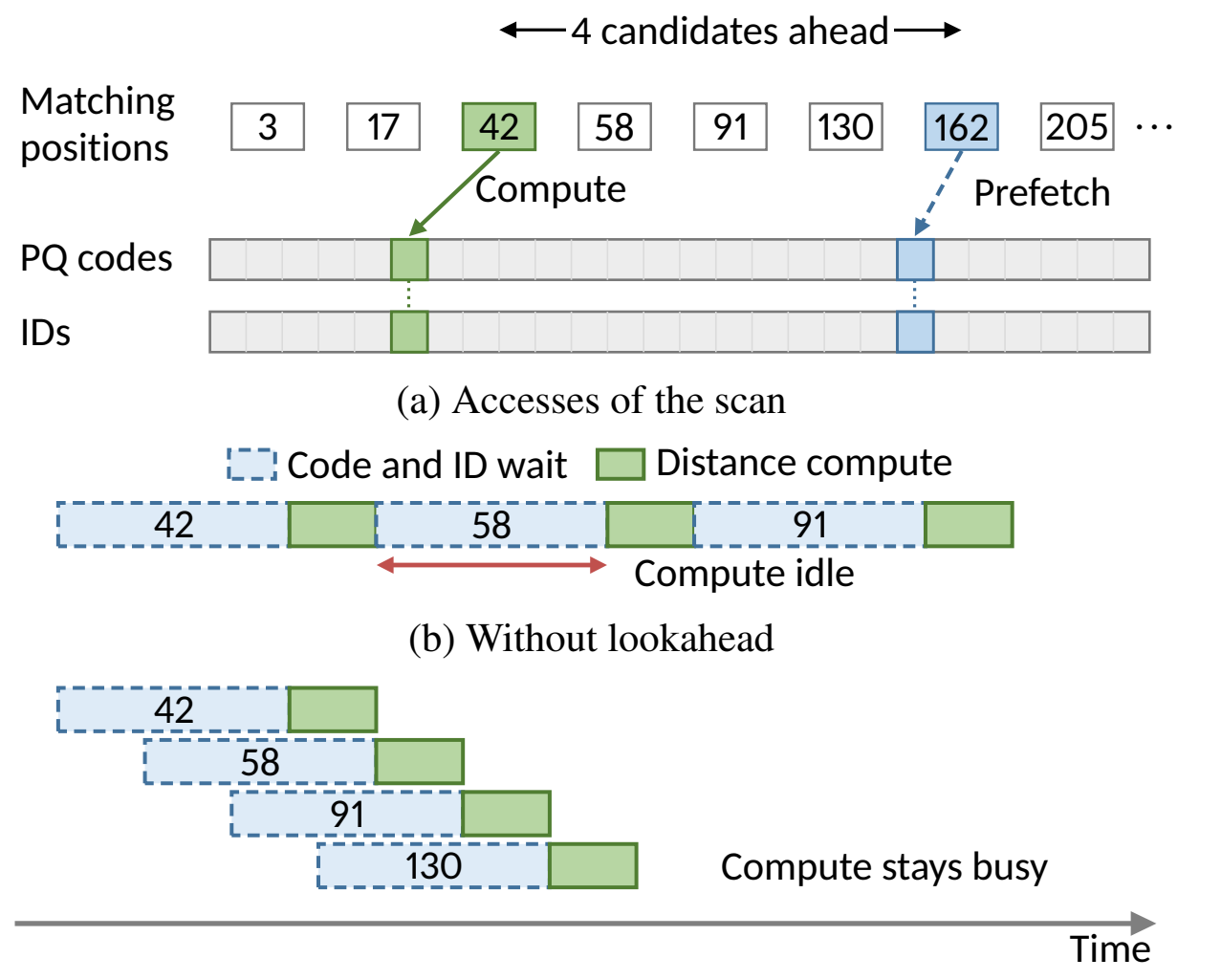

\noindent\textbf{Lookahead prefetching.}
\sys{} interleaves the memory accesses of several candidates with a fixed lookahead.
In Figure~\ref{fig:lookahead}, the lookahead is four candidates over the matching positions 3, 17, 42, 58, and so on.
Note that interleaving independent memory accesses is a known way to hide memory latency.
The lookahead distance is a machine constant chosen once to cover a round trip to DRAM.
Huge pages complement the lookahead by reducing the TLB misses on these large arrays.
Because both change only when memory is requested and how pages are mapped, the candidates and the results are identical with them enabled or disabled.

\subsection{Reranking with Bounded SSD Reads}
\label{sec:design-rerank}
\label{sec:design-d3}

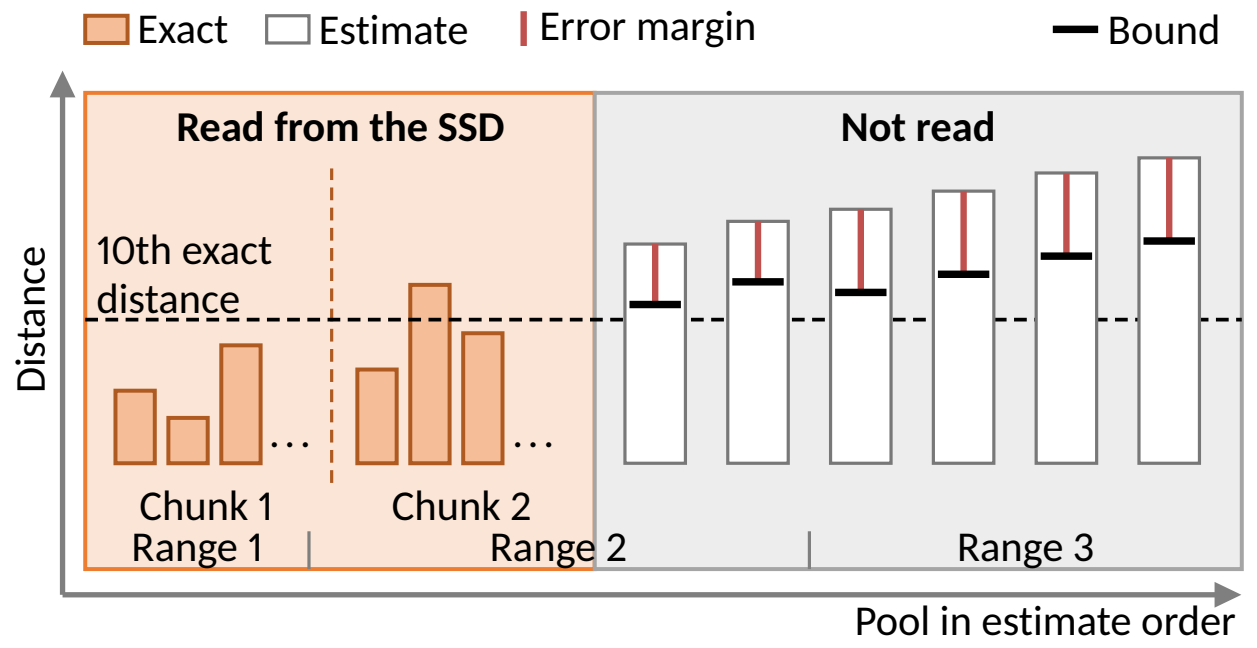
\begin{figure}[t]
  \centering
  \begin{tikzpicture}[kfig]
\def\base{0.75}
\draw[kregion={kOrange}{kOrangeL}] (0.35,0.05) rectangle (3.72,3.2);
\draw[kregion={kGray}{kGrayL}] (3.72,0.05) rectangle (8.0,3.2);
\node[khdr] at (2.04,2.98) {Read from the SSD};
\node[khdr] at (5.86,2.98) {Not read};
\draw[karr] (0.2,-0.12) -- (0.2,3.35);
\node[ktxt, rotate=90, anchor=south] at (0.12,1.7) {Distance};
\foreach \x/\h in {0.55/0.48, 0.9/0.3, 1.25/0.78, 2.15/0.62, 2.5/1.18, 2.85/0.86} {
  \draw[kpart={kOrangeS}{kOrange}] (\x,\base) rectangle ++(0.26,\h);
}
\node[ktxt] at (1.72,\base+0.12) {\dots};
\node[ktxt] at (3.33,\base+0.12) {\dots};
\draw[kOrangeS, line width=0.6pt, kdash] (1.98,0.62) -- (1.98,2.7);
\draw[black, kdash, line width=0.7pt] (0.35,\base+0.95) -- (8.0,\base+0.95);
\node[ktxt, anchor=south west, fill=kOrangeL, inner sep=0.5pt, align=left] at (0.4,\base+0.97) {10th exact\\distance};
\foreach \x/\e/\m in {3.92/1.45/0.4, 4.6/1.6/0.4, 5.28/1.68/0.55, 5.96/1.8/0.55, 6.64/1.92/0.55, 7.32/2.02/0.55} {
  \draw[kcard={kGrayS}{white}] (\x,\base) rectangle ++(0.4,\e);
  \draw[kRed, line width=1.2pt] (\x+0.2,\base+\e) -- (\x+0.2,\base+\e-\m);
  \draw[black, line width=1.4pt] (\x+0.03,\base+\e-\m) -- (\x+0.37,\base+\e-\m);
}
\node[ktxt] at (1.15,0.47) {Chunk 1};
\node[ktxt] at (2.84,0.47) {Chunk 2};
\foreach \x in {1.83,5.14} \draw[kGrayD, line width=0.6pt] (\x,0.05) -- (\x,0.3);
\node[ktxt] at (1.09,0.17) {Range 1};
\node[ktxt] at (3.48,0.17) {Range 2};
\node[ktxt] at (6.57,0.17) {Range 3};
\draw[karr] (0.2,-0.12) -- (8.0,-0.12);
\node[ktxt, anchor=north east] at (8.0,-0.16) {Pool in estimate order};
\begin{scope}[yshift=3.62cm]
  \draw[kpart={kOrangeS}{kOrange}] (0.35,-0.09) rectangle ++(0.28,0.18);  \node[ktxt, anchor=west] at (0.66,0) {Exact};
  \draw[kcard={kGrayS}{white}] (1.55,-0.09) rectangle ++(0.28,0.18);       \node[ktxt, anchor=west] at (1.86,0) {Estimate};
  \draw[kRed, line width=1.2pt] (3.25,-0.11) -- (3.25,0.12);               \node[ktxt, anchor=west] at (3.32,0) {Error margin};
  \draw[black, line width=1.4pt] (6.75,0) -- (7.05,0);                     \node[ktxt, anchor=west] at (7.08,0) {Bound};
\end{scope}
\end{tikzpicture}
  \caption{Example of calibrated stopping for 10 neighbors. Each unread candidate gets a bound (black mark), its estimate lowered by the error measured at its rank. Reading stops once every bound lies above the 10th-best exact distance read so far (dashed line), because no unread candidate is then likely to be closer.}
  \label{fig:d3}
\end{figure}

In the last step, the rerank reads the raw vectors of the best candidates from the SSD and returns the exact neighbors.
At high recall, these reads dominate query time, so \sys{} bounds them in two ways.
It stores each vector in a block-aligned slot, so each read fetches one whole vector.
It also stops each query with error bounds measured at its count, so most queries stop early.
In contrast, a vector packed across two SSD blocks costs two reads, and a fixed pool keeps reading after the answer is already complete.
Stopping once the next estimate exceeds the current $k$-th best exact distance is also unsafe, because a PQ estimate can exceed the exact distance, and a true neighbor can then lie after the stopping point.

\noindent\textbf{One aligned read per candidate.}
For example, when vectors of 3,072 bytes are packed back to back in 4\,KB blocks, half of them cross a block boundary.
Direct I/O must then read both blocks.
To avoid this, a block-aligned slot holds as many whole vectors per block as fit, and a candidate's block follows from its identifier with one division.

\noindent\textbf{Calibrated stopping.}
Multi-step search avoids missing a neighbor with a proven lower bound on the distance, but PQ codes alone give none.
Under \emph{calibrated stopping}, \sys{} instead measures the error of every estimate at each rank of the pool on the calibration queries.
The error is the exact distance minus the estimate.
For each range of pool ranks, it keeps a low quantile of these errors in a stopping table.
This quantile is usually negative.
At query time, each unread candidate receives an optimistic bound, its estimate plus the quantile of its rank range.
Reads are issued in chunks and stop once the $k$-th best exact distance is at most the smallest unread bound.

\noindent\textbf{Example.}
Figure~\ref{fig:d3} follows a query that asks for 10 neighbors.
\sys{} reads its pool in estimate order, one chunk at a time.
After each chunk, the read candidates have exact distances (orange bars), and the 10th smallest of them sets the dashed line.
\sys{} lowers the estimate of each unread candidate (white bar) by the error margin of its rank range and obtains its bound (black mark).
After Chunk 2, every unread bound lies above the dashed line, and \sys{} stops without reading the rest of the pool.

\noindent\textbf{One table per count.}
The error of the estimates at a given rank depends on the count.
A single table for all counts is therefore loose for some counts.
Our design thus keeps one table per calibration point and route.
Each plan row that stops early uses the table of its point, with a quantile level that calibration chooses for the row.
Note that sampled error quantiles are known to prune distance computations, and here a table per count decides when the SSD reads stop.
However, the quantile describes single candidates, and the unread set depends on the same estimates.
Calibrated stopping is therefore a heuristic, not a guarantee for each query.

\subsection{Updates}
\label{sec:design-updates}

Insertions and deletions should take effect without rebuilding the index or recalibrating the plan.
However, the PQ codes and the postings are packed arrays in position order, and inserting a vector into a partition in place would shift every later position.
Fortunately, the plan reads only the counts, not the positions.
We therefore keep the counts up to date but leave the packed arrays unchanged until a merge.
Instead, insertions go to a delta segment, deletions set tombstones, and an occasional merge rewrites the packed arrays.

\noindent\textbf{Delta segment and merges.}
\sys{} adds insertions to a per-partition delta segment that holds the new PQ codes, identifiers, and labels.
Because the delta segment stores no attribute values, updates currently serve label predicates only.
A new vector costs one nearest-centroid search and one encoding, because the encoder reuses the trained centroids and PQ codebooks.
During a search, the partition route scans the delta entries of a probed partition after its packed entries.
For the exhaustive route, each label keeps references to its delta entries.\looseness=-1

A deletion sets a tombstone bit, and the scan skips tombstoned vectors.
Both operations update the count of each label, which the plan reads.
Therefore, random insertions and deletions require no recalibration.
When updates are pending, an AND is counted on the packed arrays as usual, and its matching delta entries are added to the count.
A merge moves the delta segment into the packed arrays on demand and drops tombstoned vectors.
It writes each partition in ascending identifier order, which is the order of a fresh build.
A merge after inserting held-out vectors thus reproduces the arrays of an index built with them, except for vectors equally near two centroids.

\subsection{Discussion}
\label{sec:design-discussion}

\noindent\textbf{Calibration cost and workload drift.}
The cost specific to \sys{} is the calibration of the plan, because sweeping the probe count and the pool size is standard practice for a partition index.
It needs held-out queries with exact filtered answers at a few counts, runs once per workload, and yields a plan of a few kilobytes per target recall.
However, the plan needs to be recalibrated for a new target recall, for retrained centroids or codebooks, and for queries that relate to their labels differently than the calibration queries do.
When the query mix drifts, an operator may rebuild the plan offline from logged queries and replace the old one.

\noindent\textbf{Where count-calibrated reads apply.}
The principle needs two properties of an index.
First, the index must compute the count of a query before its expensive work, or at least bound it within one plan row.
Second, the recall of a fixed setting must move in one direction with the count on each route.
This second property is observed rather than guaranteed, and it fails when the matches of a label cluster near a query, which the local count detects.
A partition index has the first property for any predicate whose matches it can enumerate in memory.
It has the second for labels whose matches spread like those of the calibration labels.
A graph index lacks it, because its recall depends on the paths between the matches.
Because neither property depends on the SSD, we expect the principle to apply to in-memory partition indexes.

\begin{table}[t]
\centering
\caption{Datasets used in the evaluation.}
\label{tab:datasets}
\small
\setlength{\tabcolsep}{4pt}
\begin{tabular}{lrrlrr}
\toprule
Dataset & Vectors & Dim. & Type & Labels & Labels/vec \\
\midrule
LAION-100M & 100M & 768 & float & 10,824 & 5.0 avg \\
SIFT-1B & 1B & 128 & uint8 & 43,592 & 6.0 avg \\
\midrule
YFCC-10M & 10M & 192 & uint8 & 200,386 & 10.8 avg \\
YT-5M & 5M & 1024 & float & 3,862 & 3.0 avg \\
\midrule
PubMed & 23.9M & 1024 & float & 29,912 & 12.0 avg \\
Amazon & 35.1M & 768 & float & 12,996 & 6.0 avg \\
\bottomrule
\end{tabular}
\end{table}

\begin{figure*}[t!]
  \centering
  \includegraphics[width=0.95\linewidth]{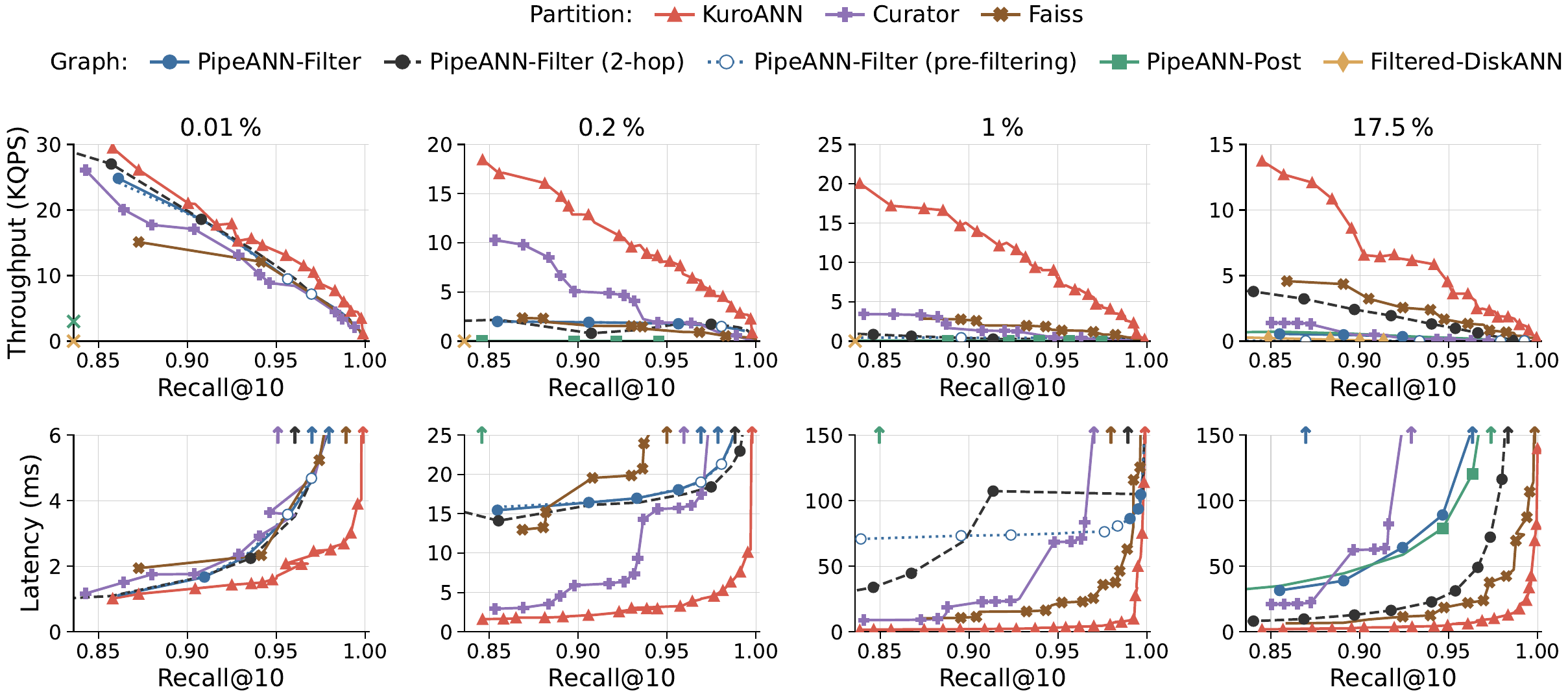}
  \caption{Throughput and latency against recall. \sys{} leads every baseline, and only \pf{} comes close at the rarest selectivity.}
  \label{fig:curves}
\end{figure*}

\section{Implementation}
\label{sec:impl}

We implement the \sys{} search engine in about 1.5K lines of C++ on \texttt{liburing}.
Python and C++ tools build the index and calibrate the plans.
The builder trains the centroids and the PQ codebooks with \faiss~\cite{faiss}.
Each PQ code holds one byte per piece, which gives 64 bytes on LAION-100M and YFCC-10M, 32 on SIFT-1B, and 128 on YT-5M.
The number of partitions is the power of two nearest to one partition per 2{,}500 vectors, at most 32{,}768.
An HNSW graph~\cite{hnsw} over the centroids finds the nearest partitions.
The scan prefetches eight candidates ahead~\cite{chenpref}, and the PQ codes and postings are backed by transparent huge pages~\cite{ingens}.
Each search thread issues asynchronous reads through its own \texttt{io\_uring} ring~\cite{iopassthru}.
A fixed row submits its whole pool at once to keep many reads outstanding~\cite{unwrittenssd}.
In contrast, a row with the calibrated stop submits 32 reads at a time and checks the stopping rule after each batch.

\section{Evaluation}
\label{sec:eval}

\subsection{Methodology}
\label{sec:eval-method}

\noindent\textbf{Testbed.}
Our server has two Intel Xeon Silver 4514Y CPUs with 16 cores each, 256\,GB of DDR5 memory, and an 8\,TB WD Black SN850X NVMe SSD.
The SSD serves 872K random 4\,KiB reads per second.
The server runs Ubuntu~24.04 with Linux~6.17.
Every system reads its on-SSD files with \texttt{O\_DIRECT} after the page cache is released.
Each system searches with 32 threads in a closed loop~\cite{openclosed}, pinned to the 32 physical cores, and interleaves memory across the two sockets.

\noindent\textbf{Compared systems.}
We compare \sys{} with three graph-based systems.
\pf{}~\cite{pipeannfilter} filters speculatively, \pfpost{} forces the post-filtering path of \pipeann{}~\cite{pipeann,pipeannfilter}, and \fdiskann{}~\cite{fdiskann} builds the labels into the graph.
We report the faster \pf{} build with and without its 2-hop neighbors, and also with its pre-filtering path forced.
However, the 2-hop neighbors do not fit on the SSD for SIFT-1B and for the random labels of \S\ref{sec:eval-deep}.
\fdiskann{} runs the filtered SSD workflow of DiskANN with two crash fixes that leave its search unchanged, and it reaches recall 0.99 on LAION-10M with uniform labels.

We also adapt the in-memory partition indexes \curator{}~\cite{curator26} and \faiss{}~\cite{faiss} to the SSD.
Both rank the PQ codes of \sys{} and rerank from its on-SSD store, and \faiss{} also uses its centroids.
On YFCC-10M, the adapted \curator{} reaches the recall of the official in-memory \curator{} at the same search budget.
Below a count threshold, \faiss{} scans the matches as its authors suggest~\cite{faiss}, and above it probes the nearest partitions.

\noindent\textbf{Datasets.}
Table~\ref{tab:datasets} lists the six datasets.
LAION-100M~\cite{laion5b,vdbb} and SIFT-1B~\cite{bigann11} have no natural labels for filtered search.
We therefore give them synthetic labels, five and six per vector, from k-means clusters of several sizes.
Synthetic labels let every band reach a chosen selectivity, and labels from clusters resemble real categories, whose members tend to be similar.
The bands range from 0.01\% to 17.5\% on LAION-100M and to 16.6\% on SIFT-1B.
Because these labels place the vectors of a label close together, \S\ref{sec:eval-deep} also evaluates labels drawn independently of the vectors.
YFCC-10M and YT-5M have natural labels.
As in the evaluation of \pf{}~\cite{pipeannfilter}, we form AND queries on YFCC-10M and OR queries on YT-5M, and we group each into four bands.
PubMed and Amazon come from two real applications with natural filters (\S\ref{sec:eval-apps}).

\noindent\textbf{Calibration and comparison.}
\sys{} calibrates its plans in 4.2\,h on 500 held-out queries per band of the evaluation labels.
It then freezes them before the evaluation, which uses another 500 queries per band.
Each baseline instead sweeps its parameters on the evaluation queries.
The plans meet every target on every dataset.
Each baseline is read at the recall of \sys{}, interpolating between two settings when none matches it.
Each query asks for 10 neighbors, and ties at the 10th distance count as found.
Throughput in queries per second (QPS) and latency come from timed passes.

\subsection{Main Results}
\label{sec:eval-main}

\noindent\textbf{Throughput and latency against recall.}
Figure~\ref{fig:curves} shows the throughput and median latency against recall at four bands of LAION-100M. A cross marks a point below the plotted range, and an arrow marks one above it.
\fdiskann{} appears only in the throughput row because it reports no median latency.
We can see that \sys{} leads every baseline up to recall 0.999 at every band except 0.01\%.
There, \pf{} takes its pre-filtering path, which ranks only the few matching PQ codes, and the two curves nearly coincide below recall 0.93.
At the other bands, its graph traversal spends SSD reads on detours.
Forcing its pre-filtering path (dotted) avoids these reads but ranks every matching PQ code.
Consequently, at 1\% and the 0.9 target, \sys{} still answers 32.0$\times$ faster at the median than the faster of the two paths.
In addition, \curator{} and \faiss{} stay below \sys{} at all four bands.

\begin{figure}[t]
  \centering
  \includegraphics[width=0.85\linewidth]{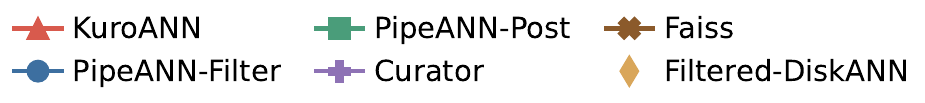}\\[1pt]
  \subfloat[Target recall 0.9]{\includegraphics[width=0.46\linewidth]{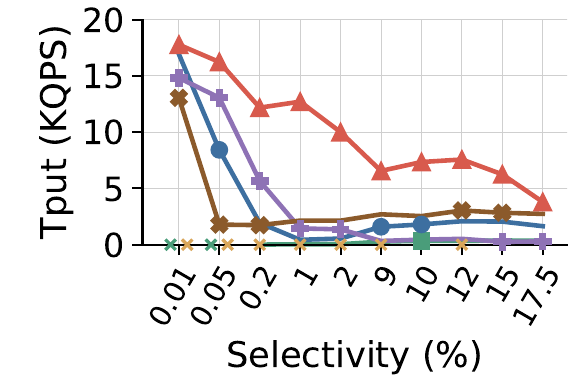}}\hfill
  \subfloat[Target recall 0.99]{\includegraphics[width=0.46\linewidth]{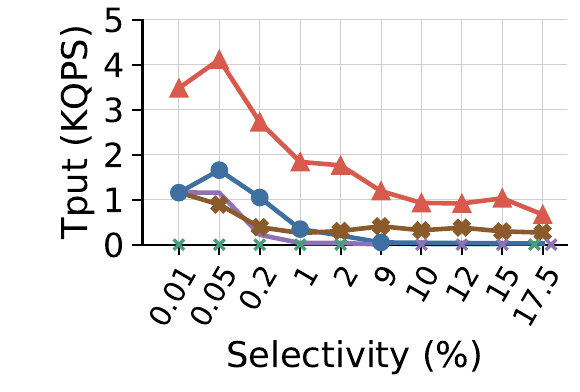}}
  \caption{Impact of selectivity. \sys{} outperforms every baseline at every selectivity.}
  \label{fig:frontier}
\end{figure}

\begin{figure}[!t]
  \centering
  \includegraphics[width=0.95\linewidth]{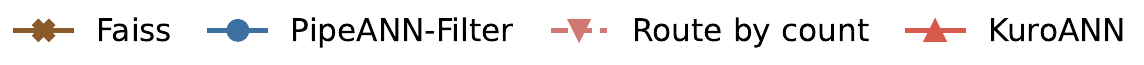}\\[1pt]
  \subfloat[Throughput]{\includegraphics[width=0.48\linewidth]{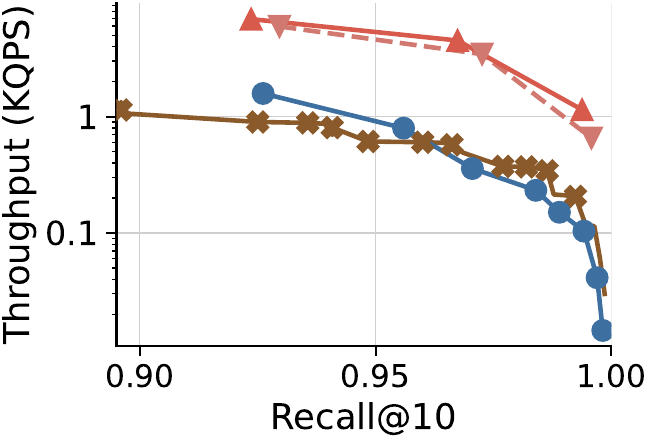}}\hfill
  \subfloat[Tail latency]{\includegraphics[width=0.48\linewidth]{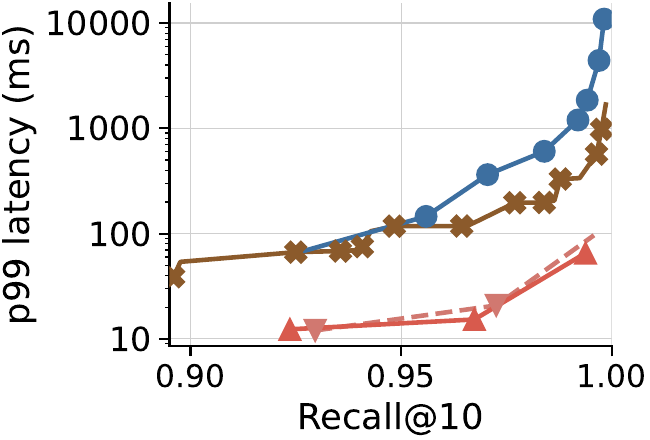}}
  \caption{Impact of mixed selectivities. With one configuration for the whole stream, every baseline falls far behind \sys{} in throughput and tail latency.}
  \label{fig:mix}
\end{figure}

\noindent\textbf{Impact of selectivity.}
Figure~\ref{fig:frontier} shows that \sys{} leads every baseline at every band at the 0.9 and 0.99 targets of LAION-100M.
A cross marks a baseline below the recall of \sys{}.
At 0.99, \sys{} outperforms \pf{} by 10.0$\times$ in geometric mean and by up to 34.5$\times$, with the largest leads from 9\% upward.
This is because \pf{} reaches the recall of \sys{} there only on its pre-filtering path, which ranks millions of matching PQ codes per query.
The lead is smaller at the three rarest bands, where \pf{} reads 2.11$\times$ to 3.54$\times$ more vectors per query than \sys{}.
\sys{} also outperforms \faiss{} by up to 9.10$\times$ at 0.9, because the fixed pool of \faiss{} must cover the queries of a band that need the most reads.
Meanwhile, \curator{} misses the recall of \sys{} above 9\% at 0.99, and \fdiskann{} reaches it only at three bands at 0.9.
\sys{} outperforms \pfpost{} by 11$\times$ or more, because filtering disconnects its graph or forces long detours.

\begin{figure}[t]
  \centering
  \subfloat[Speedup by band]{\includegraphics[width=0.46\linewidth]{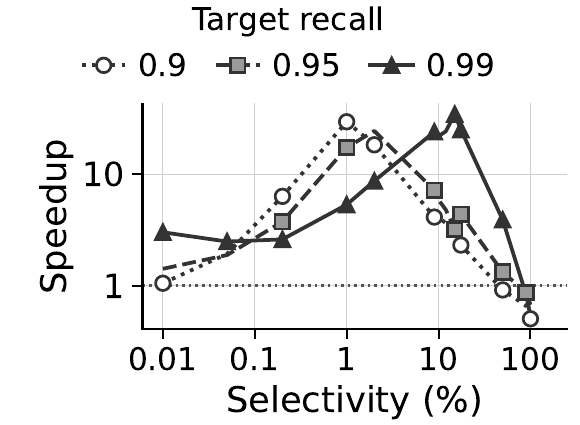}}\hfill
  \subfloat[100\% (no filter)]{\includegraphics[width=0.46\linewidth]{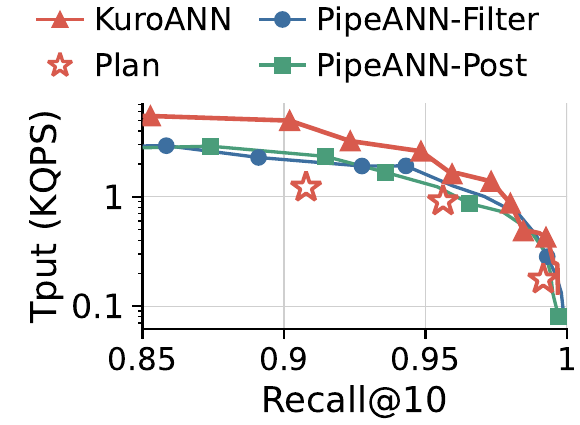}}
  \caption{Impact of broad predicates. The lead of \sys{} peaks at the middle selectivities and shrinks at both ends, yet \sys{} still leads without a filter.}
  \label{fig:broad}
\end{figure}

\noindent\textbf{Impact of mixed selectivities.}
Figure~\ref{fig:mix} sends a stream of 500 LAION-100M queries, 50 from each band, to every system.
Each system runs its best single configuration on the whole stream.
The dashed line runs the engine of \sys{} with the threshold rule of \faiss{}, which we call route by count, instead of its plans.
We can see that \sys{} is 6.9$\times$ to 8.2$\times$ faster than \faiss{} and 4.3$\times$ to 11.1$\times$ faster than \pf{} at the recall of its three plans.
It also has better p99 latency than both by 5.4$\times$ to 28.4$\times$.
This is because the counts in the stream lie three orders of magnitude apart, and a single setting must serve them all.

\noindent\textbf{Impact of broad predicates.}
Predicates that most vectors satisfy approach unfiltered search, where a graph index needs no repair.
We therefore run OR predicates that 49.9\% and 89.7\% of LAION-100M satisfy and one that covers every vector, with the plans frozen from the bands up to 17.5\%.
Figure~\ref{fig:broad}~(a) shows that the lead over the faster of \pf{} and \pfpost{} peaks at the middle bands, at up to 34.5$\times$ at 0.99, and shrinks at both ends.
Without a filter, \pf{} is up to 1.85$\times$ faster than the plans, because they probe more partitions than these predicates need.
However, a fixed setting of \sys{} still leads the faster of \pf{} and \pfpost{} by 1.17$\times$ to 1.99$\times$, as Figure~\ref{fig:broad}~(b) shows (stars mark the plans).

\begin{figure}[t]
  \centering
  \includegraphics[scale=0.4]{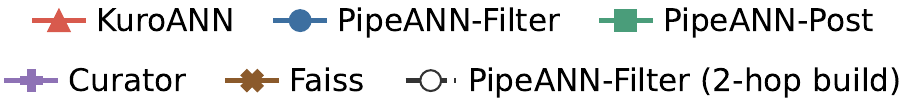}\\[1pt]
  \subfloat[YFCC-10M (AND)]{\includegraphics[width=0.46\linewidth]{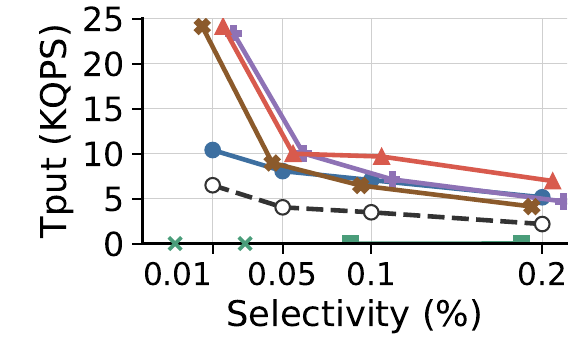}}\hfill
  \subfloat[YT-5M (OR)]{\includegraphics[width=0.46\linewidth]{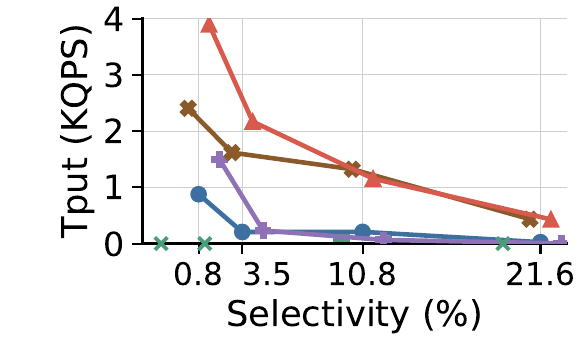}}
  \caption{Impact of AND and OR predicates. \sys{} leads \pf{} on both, and the adapted partition indexes come close.}
  \label{fig:andor}
\end{figure}

\noindent\textbf{Impact of AND and OR predicates.}
Figure~\ref{fig:andor} reports the AND queries on YFCC-10M, which match below 0.5\% of the vectors.
It also reports the OR queries on YT-5M, which match 0.8\% to 21.6\%. The dashed line shows \pf{} with its 2-hop neighbors, and a cross marks a setting below the recall of \sys{}. We skip \fdiskann{}, whose search takes one label per query.
On the AND queries, \sys{} leads \pf{} by 1.24$\times$ to 2.32$\times$, although \pf{} also pre-filters there.
This is because \pf{} reads the label lists from the SSD and two 4\,KiB sectors for each candidate that it reranks.
\sys{} instead intersects postings in memory and reranks 16 to 32 candidates.
Because these counts are small, \curator{} and \faiss{} also rank few PQ codes, and one of them ties \sys{} at each of the two rarest bands.
On the OR queries, \sys{} leads \pf{} by up to 17.3$\times$.
However, \faiss{}, the strongest baseline, is up to 1.15$\times$ faster at two of the four bands and up to 1.62$\times$ slower at the other two.
At exactly the 0.99 target rather than at the recall of \sys{}, \faiss{} is faster at three of the four bands.

\begin{figure}[t]
  \centering
  \includegraphics[scale=0.4]{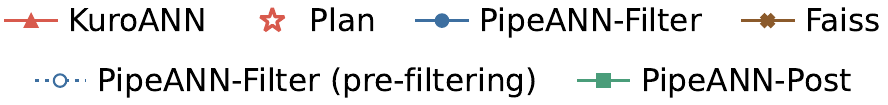}\\[1pt]
  \subfloat[0.01\%]{\includegraphics[width=0.46\linewidth]{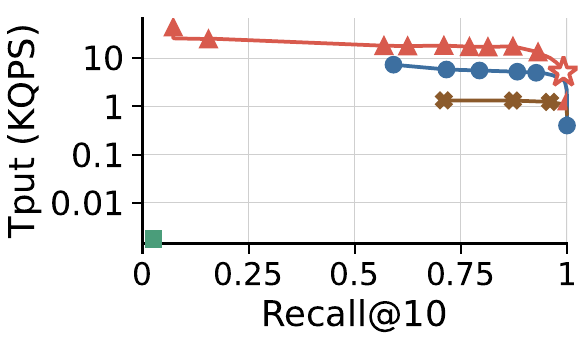}}\hfill
  \subfloat[1\%]{\includegraphics[width=0.46\linewidth]{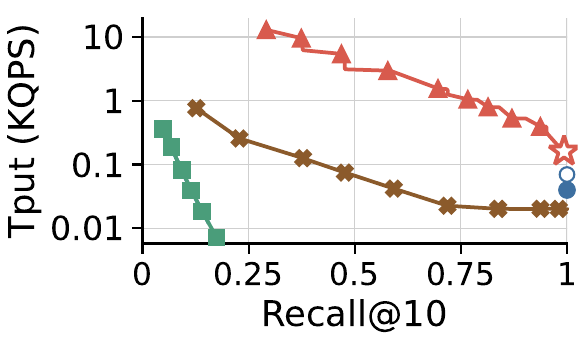}}
  \caption{Throughput against recall on the 1B dataset (SIFT-1B). \sys{} keeps its lead at 1B vectors, and \pf{} comes closest at the rarest selectivity. Stars mark the plans at the 0.99 target.}
  \label{fig:1bcurves}
\end{figure}

\noindent\textbf{Impact of dataset scale.}
Figure~\ref{fig:1bcurves} compares the recall-throughput curves at the 0.01\% and 1\% bands of SIFT-1B.
We skip \curator{}, which would exceed the memory, and \fdiskann{}, which stays below recall 0.03 at these bands on LAION-100M.
We read \pf{} at its fastest path, which is pre-filtering at both bands and reaches every target.
At 0.01\%, \pf{} pre-filters on its own, so the forced path appears only at 1\%.
At 0.99, \pf{} comes closest at 0.01\%, at 1.48$\times$ slower, because only 100K vectors match there and pre-filtering searches them directly.
At 1\%, \sys{} is 2.11$\times$ faster, and \faiss{} is 4.6$\times$ to 8.0$\times$ slower than \sys{} at the two bands.

\begin{figure}[t]
  \centering
  \includegraphics[width=0.95\linewidth]{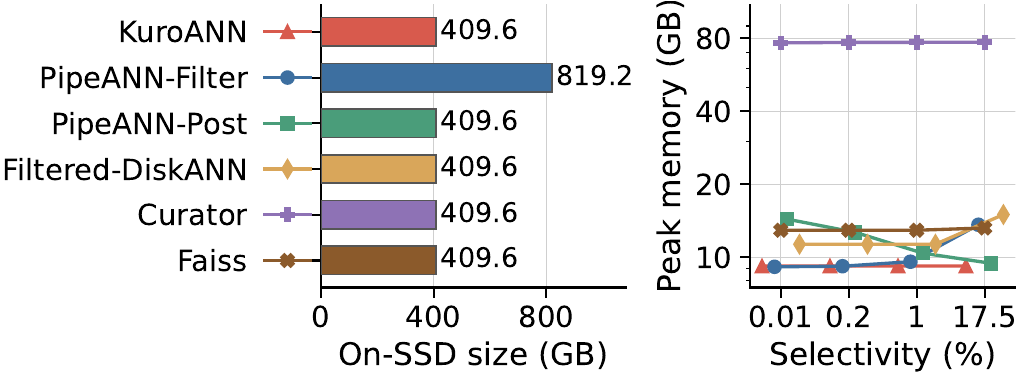}
  \caption{SSD and memory footprint. \sys{} stores much less than \pf{} with 2-hop neighbors and uses the least memory except at the two sparsest bands, where \pf{} uses about as much.}
  \label{fig:footprint}
\end{figure}

\noindent\textbf{SSD and memory footprint.}
Figure~\ref{fig:footprint} shows the on-SSD size and the peak memory of each system on LAION-100M.
\sys{} serves every band from a 409.6\,GB store, half of the 819.2\,GB index of \pf{} with 2-hop neighbors.
In memory, \sys{} uses at most 9.2\,GB, about as much as \pf{} at the 0.01\% and 0.2\% bands, less than \pf{} at the broader bands, and less than \fdiskann{} at every band, because all of them keep the 64-byte PQ codes of every vector.
Its postings and position-to-identifier array add only 24 bytes per vector.
In contrast, \faiss{} uses at most 13.2\,GB, because it also keeps the labels of every vector and a map from each vector to its PQ code.
\curator{} holds 77.0\,GB, because each tree node carries a Bloom filter for all labels.

\subsection{Applications}
\label{sec:eval-apps}

We now show that \sys{} speeds up two real applications of filtered vector search when their queries arrive at a rising rate.
These experiments run on a second server with an Intel Xeon Gold 5512U CPU with 28 cores, 128\,GB of memory, and an 800\,GB Dell DC NVMe 7450 SSD, and each system searches with 28 threads.
Each system receives the evaluation queries of a workload as Poisson arrivals, and we report the p99 latency from arrival to completion at the 0.99 target. A line ends at the first rate whose p99 latency exceeds 10\,s.
\sys{} runs its 0.99 plans, and each baseline uses, per band, its fastest setting that reaches the recall of \sys{}, or its most accurate one where none does, a choice that favors the baselines.

\begin{figure}[t]
  \centering
  \includegraphics[scale=0.4]{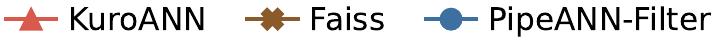}\\[1pt]
  \subfloat[RAG retrieval on PubMed]{\includegraphics[width=0.48\linewidth]{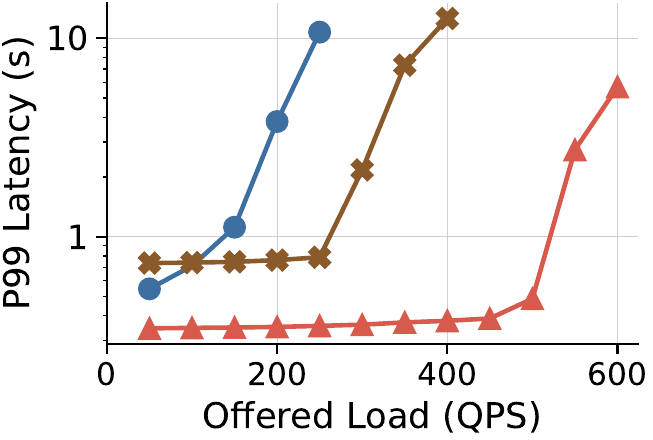}}\hfill
  \subfloat[Recommendation on Amazon]{\includegraphics[width=0.48\linewidth]{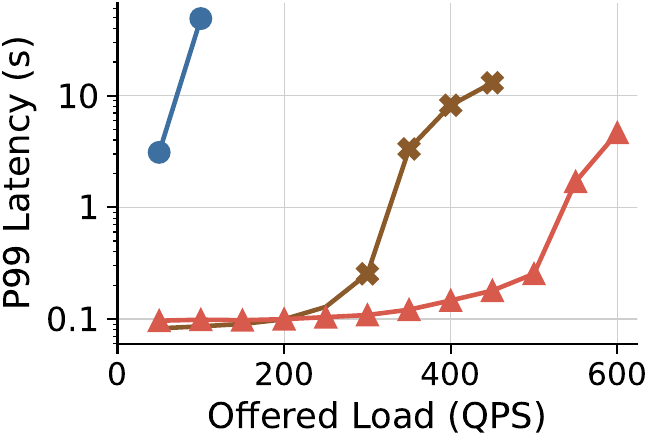}}
  \caption{Applications under load. \sys{} keeps a low tail latency at loads where the baselines saturate.}
  \label{fig:load}
\end{figure}

\begin{figure}[t]
  \centering
  \subfloat[Policies against the plan]{\begin{minipage}[b]{0.61\linewidth}\centering\includegraphics[scale=0.44]{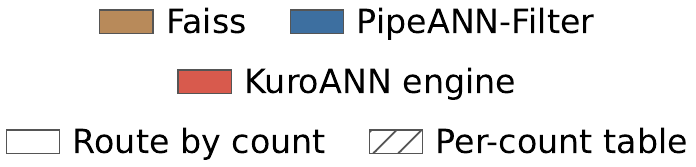}\\[1pt]\includegraphics[width=\linewidth]{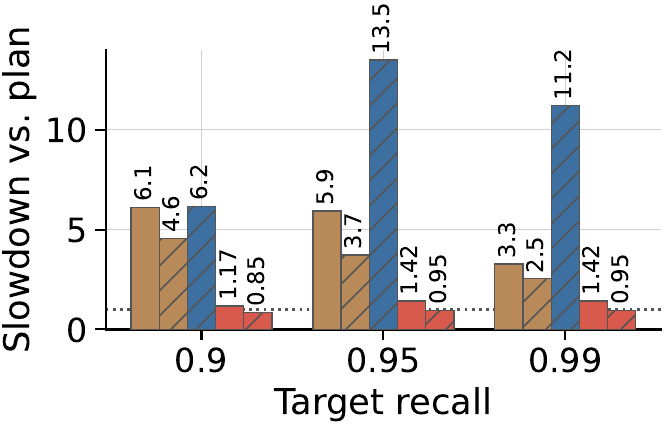}\end{minipage}}\hfill
  \subfloat[Postings over a bitmap, no prefetching]{\begin{minipage}[b]{0.37\linewidth}\centering\includegraphics[scale=0.44]{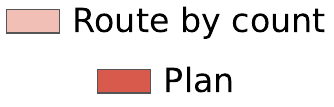}\\[1pt]\includegraphics[width=\linewidth]{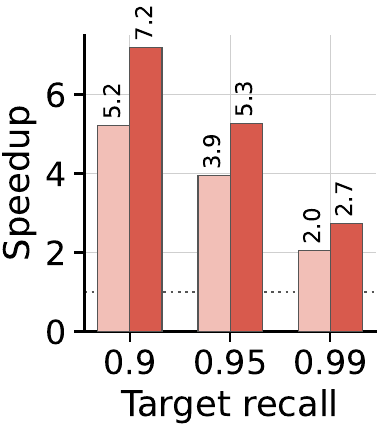}\end{minipage}}
  \caption{Where the lead comes from. The faster work in memory and the reads that the count sets multiply.}
  \label{fig:k50}
\end{figure}

\noindent\textbf{Retrieval for RAG on PubMed.}
A RAG system retrieves the abstracts closest to a question, often within a scope such as a publication type or a range of years.
PubMed~\cite{pubmedmv} holds the BGE-M3~\cite{bgem3} embeddings of 23.9 million abstracts, filtered by publication type, MeSH term, and year, and its queries are bioRxiv titles.
Figure~\ref{fig:load}~(a) shows that \sys{} keeps its p99 latency below 0.5\,s up to 500 QPS, whereas \faiss{} saturates at 274 QPS and \pf{} at 180 QPS.
This is because \sys{} serves this stream 1.6$\times$ and 2.4$\times$ faster than \faiss{} and \pf{}, and a queue forms only once the offered load exceeds the throughput of a system.

\noindent\textbf{Recommendation on Amazon.}
A store recommends the items closest to what a shopper bought, within the category, rating, and price that the shopper picks.
Amazon holds the BLaIR~\cite{blair} embeddings of 35.1 million items of Amazon Reviews'23, filtered by category, rating, and price, and each query is the mean of a shopper's recent items.
Figure~\ref{fig:load}~(b) shows that \sys{} serves 492 QPS at a p99 of 254\,ms, while \faiss{} saturates at 310 QPS and \pf{} at 58 QPS.
\sys{} serves this stream 1.6$\times$ and 8.6$\times$ faster than \faiss{} and \pf{}.
Below 200 QPS, however, \faiss{} has a p99 latency similar to that of \sys{}.

\subsection{Deep Dive}
\label{sec:eval-deep}

\noindent\textbf{Where the lead comes from.}
Figure~\ref{fig:k50} composes the mixed stream from per-band runs and splits the lead over \faiss{} into two parts.
The composition also lets every policy use a per-count table chosen on the stream (hatched).
The first part is the work in memory, where the engine of \sys{} is faster than that of \faiss{} under the same route by count at every target, most at 0.9.
This is because it touches only the matches in the probed partitions.
The second part is the reads that the count sets.
Figure~\ref{fig:k50}~(a) shows that the plan runs 1.17$\times$ to 1.42$\times$ faster than route by count, because each route over-provisions one budget for the band that needs the most reads.
Note that a per-count table changes the settings of a baseline but not its work per setting.
Therefore, the plan still outperforms \faiss{} and \pf{} by more than 2$\times$ when they use one.
Figure~\ref{fig:k50}~(b) shows that the two parts multiply, because the postings gain more under the plan than under route by count.

\begin{figure}[t]
  \centering
  \includegraphics[width=0.8\linewidth]{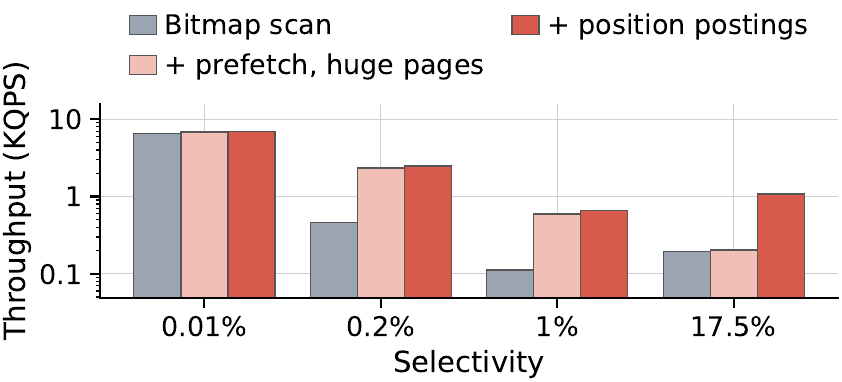}
  \caption{Cumulative ablation of the PQ code scan on LAION-100M at one fixed setting (512 probed partitions, pool of 128). Every step returns the same results.}
  \label{fig:ablation}
\end{figure}

\noindent\textbf{Impact of the scan mechanisms.}
Figure~\ref{fig:ablation} adds the mechanisms of the scan one at a time at a scan-heavy fixed setting.
Lookahead prefetching with huge pages raises the throughput by about 5$\times$ at 0.2\% and 1\%, because without it the scan waits on memory at almost every candidate.
At 17.5\%, the position-ordered postings raise the throughput by 5.31$\times$ instead, because the scan then touches only the matching codes of the probed partitions.
At 0.01\%, where the query scans few codes, the two steps add only 5\%.

\begin{figure}[t]
  \centering
  \includegraphics[scale=0.4]{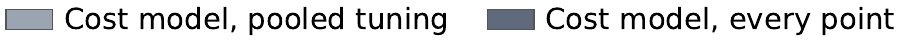}\\[1pt]
  \includegraphics[width=0.8\linewidth]{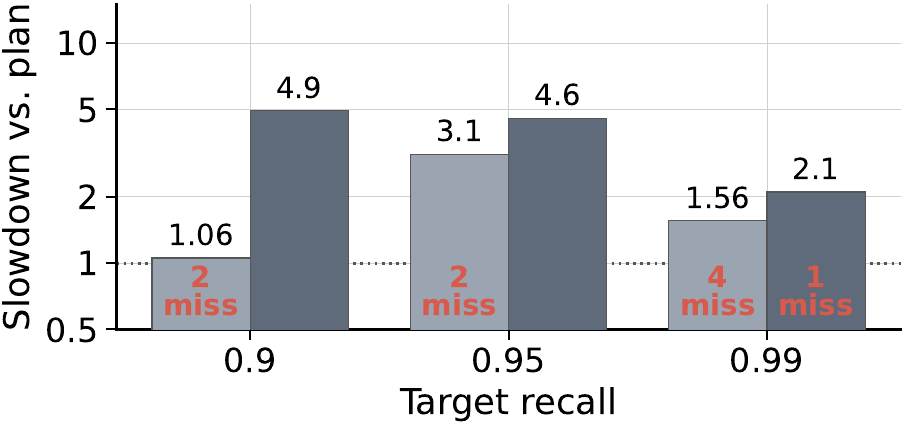}
  \caption{Cost-model plans. A number in a bar counts the bands, of 10, below the target.}
  \label{fig:costplan}
\end{figure}

\noindent\textbf{A cost model in place of measured rows.}
A database optimizer would derive the plan from a cost model, so we also build such plans on the engine of \sys{}.
Each takes the route that scans fewer codes, probes the fewest partitions expected to hold a fixed number of matches, and reads one pool per route without stopping early.
Figure~\ref{fig:costplan} compares them with the plan on the stream of Figure~\ref{fig:k50}.
With constants tuned for the recall pooled over the calibration points, they leave two to four bands below each target and run 1.06$\times$ to 3.13$\times$ slower.
Tuned for every point, they miss no more bands than the plan but run 2.11$\times$ to 4.94$\times$ slower.
This is because one rule for the probes and one pool per route cannot follow how the recall of a budget changes with the count.

\begin{figure}[t]
  \centering
  \subfloat[QPS]{\includegraphics[width=0.48\linewidth]{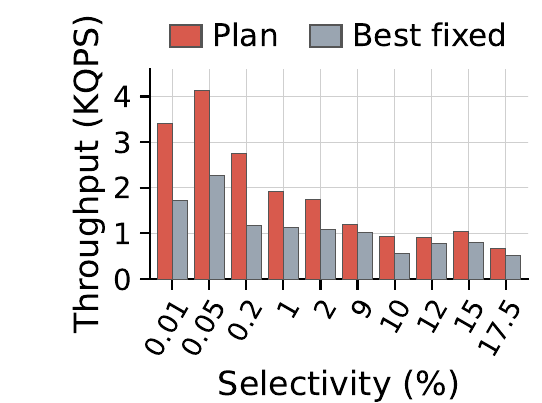}}\hfill
  \subfloat[SSD reads per query]{\includegraphics[width=0.48\linewidth]{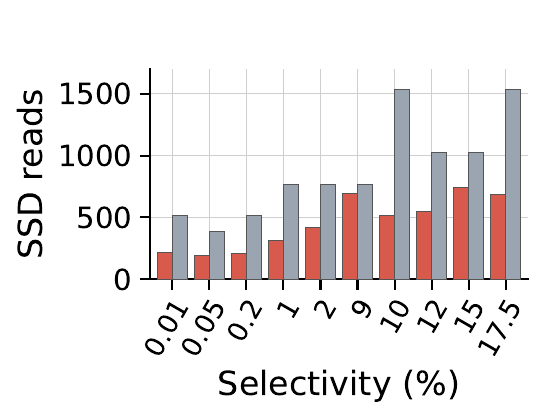}}\\[1pt]
  \subfloat[Tables rebuilt on disjoint labels]{\includegraphics[width=0.48\linewidth]{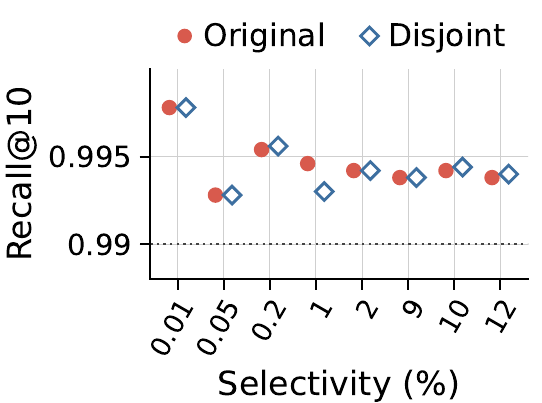}}\hfill
  \subfloat[One table per route]{\includegraphics[width=0.48\linewidth]{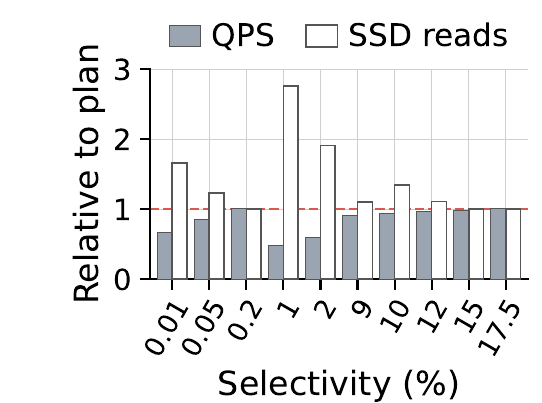}}
  \caption{Impact of calibrated stopping. Stopping by the count makes the plan faster than the best fixed budget, and a stopping table transfers to other labels but not to other counts.}
  \label{fig:stopping}
\end{figure}

\noindent\textbf{Impact of calibrated stopping.}
Figure~\ref{fig:stopping}~(a, b) compares the 0.99 plan with the best fixed budget per band on the same engine, the policy of AnalyticDB-V~\cite{adbv} applied with exact counts.
The plan is 1.18$\times$ to 2.33$\times$ faster, because most queries stop early and read 1.10$\times$ to 2.97$\times$ fewer vectors.
The gain thus comes from the count of each query, not from tuning, as the fixed budget is the best one for each band.
Figure~\ref{fig:stopping}~(c) rebuilds each stopping table on a disjoint label of the same count, which eight bands have, and recall changes by at most 0.0016.
This is because a table calibrates the estimate error of the PQ codes, which depends little on which label selects them.
However, a table does depend on the count.
Figure~\ref{fig:stopping}~(d) gives every row of a route the table of a single count.
The plan then reads up to 2.8$\times$ more vectors and runs up to 2.06$\times$ slower, while recall rises by at most 0.0016.
This is because how far an estimate errs at a given pool rank changes with the count.

\begin{figure}[t]
  \centering
  \includegraphics[width=0.75\linewidth]{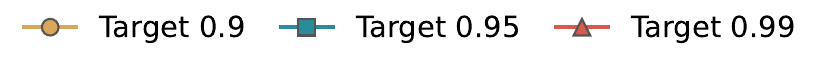}\\[1pt]
  \subfloat[Recall above the target]{\includegraphics[width=0.48\linewidth]{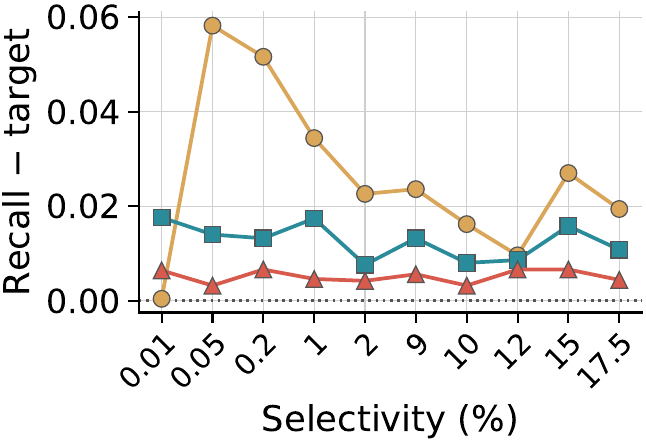}}\hfill
  \subfloat[QPS relative to the frozen plan]{\includegraphics[width=0.48\linewidth]{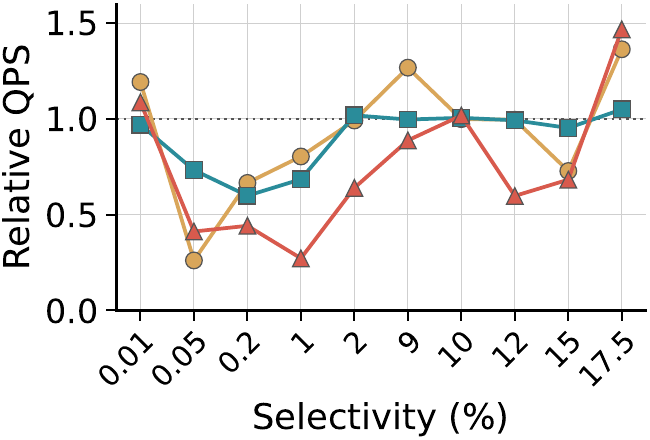}}
  \caption{Bands that calibration has not seen. The plans still meet every target, mostly at lower throughput than the frozen plans.}
  \label{fig:lobo}
\end{figure}

\noindent\textbf{Bands that calibration has not seen.}
Figure~\ref{fig:lobo} serves each band with a plan calibrated without that band, whose count then falls between the remaining calibration points or beyond the last one.
These plans meet every target at every band.
However, the frozen plans are up to 3.83$\times$ faster.
This is because a fixed setting that both neighboring points validate serves a count between them, without a stopping rule.
Therefore, a deployment needs calibration points near its common counts.

\begin{figure}[t]
  \centering
  \subfloat[Speedup over each baseline]{\includegraphics[width=0.95\linewidth]{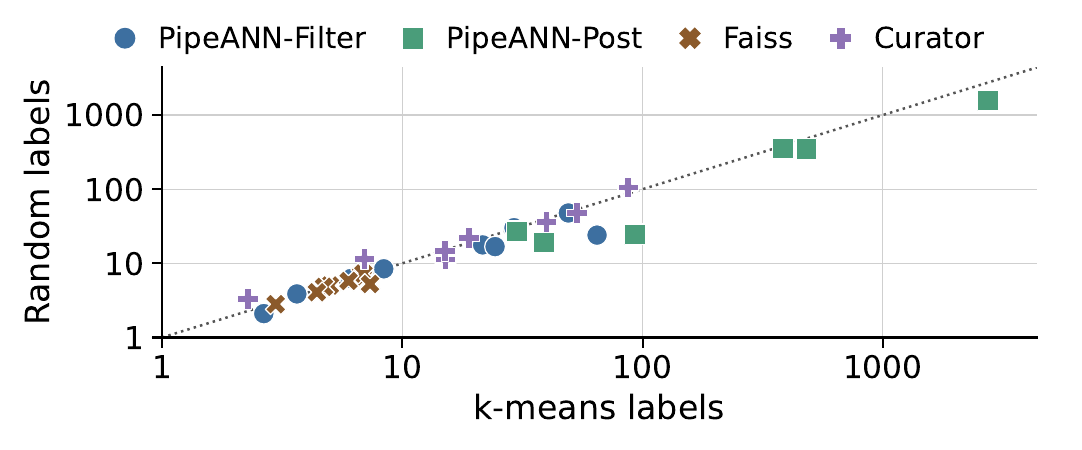}}\\[2pt]
  \includegraphics[scale=0.4]{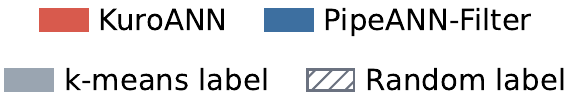}\\[1pt]
  \subfloat[SSD reads per query at the 0.99 target]{\includegraphics[width=0.85\linewidth]{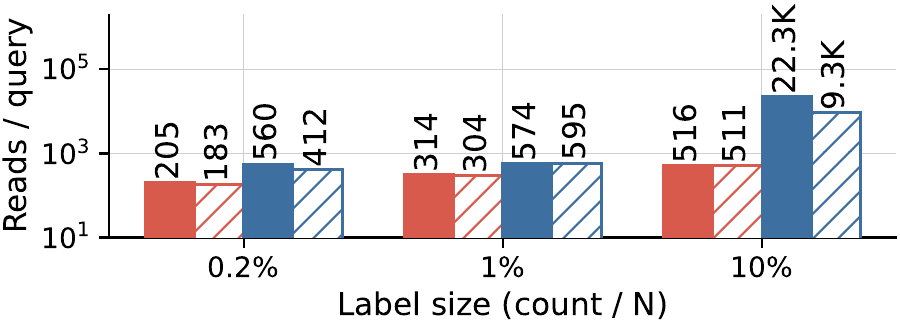}}
  \caption{Labels independent of the vectors. \sys{} leads every baseline on random labels as well, and its reads follow the count, not how the labels are drawn.}
  \label{fig:randlabel}
\end{figure}

\noindent\textbf{Labels independent of the vectors.}
Because the k-means labels could favor a partition index, we also draw labels uniformly at random with the sizes of the 0.2\%, 1\%, and 10\% labels.
The frozen plans meet every target on both kinds of label, and Figure~\ref{fig:randlabel}~(a) shows that \sys{} leads every baseline on both.
Figure~\ref{fig:randlabel}~(b) shows that at equal count \sys{} reads about as many vectors for both kinds, while \pf{} reads up to 2.4$\times$ more for one kind.
This is because the work of a partition index follows the count, while that of a graph depends on how the matches connect.

\section{Related Work}
\label{sec:related}

\noindent\textbf{Vector search on SSDs.}
Most on-SSD systems search a graph stored with the raw vectors~\cite{diskann,starling,pipeann,odinann}.
Partition indexes on SSDs read whole partitions~\cite{spann,spfresh,helmsman} or rerank compressed candidates~\cite{zoom,grip,fusionanns,rabitqssd}, all without a filter.

\noindent\textbf{Filtered vector search.}
On SSDs, graph systems post-filter results or densify the graph~\cite{pipeannfilter}, build the labels into the graph~\cite{fdiskann}, or add a flat scan of the matches~\cite{cosmosdb}.
A recent system, NemoANN~\cite{nemoann}, keeps a truncated copy of every neighbor list in memory, but its memory overhead is large.
Their work depends on the paths between the matches.
In memory, filtered graphs densify a predicate-agnostic graph~\cite{acorn} or add filter-aware edges and subindexes~\cite{sieve,tagfilter,serf,irangegraph,unify,nhq,vecflow}.
VBASE~\cite{vbase} and E2E~\cite{e2e} stop a filtered traversal by a relaxed-monotonicity check or a predicted budget of distance computations.
\curator~\cite{curator26} keeps sorted lists of each label's matches in a clustering tree, which our postings apply to flat partitions.
IVF$^2$~\cite{ivf2}, UBASE~\cite{ubase}, and vector engines~\cite{milvus,qdrantplanning,weaviatefiltering} choose a route by the count or a cost estimate.
Similarly, Pinecone~\cite{pinecone}, AnalyticDB-V~\cite{adbv}, and Lu et al.~\cite{lufvs26} relate the search budget to the selectivity as database optimizers choose plans by selectivity~\cite{pqo,vaidya2021pqo}.

\noindent\textbf{Search effort per query.}
Quake~\cite{quake} sets each query's probes from a geometric model and only sketches scaling it by the matches per partition.
Helmsman~\cite{helmsman} predicts the probes with a learned model, and Ada-ef~\cite{adaef} looks up a graph's search width in an offline table.
In addition, LAET~\cite{earlyterm}, DARTH~\cite{darth}, and Auncel~\cite{auncel} stop at the point that a model predicts for the target recall.
Multi-step search~\cite{multistepknn}, RaBitQ~\cite{rabitq,rabitqssd}, TRIM~\cite{trim}, and DADE~\cite{dade} stop or skip reranking by a distance bound or by sampled error quantiles.
All of them work without a filter.
\sys{} instead sets the reads of each query from its count and a target recall, including when it stops.

\section{Conclusion}
\label{sec:conclusion}
We presented \sys, a filtered vector search engine for SSDs that avoids unnecessary SSD reads at high recall.
The key idea of \sys is to set how widely each query searches and when it stops reading from its match count, known before the first read.
Its work in memory touches only the matches.
We hope that this work can contribute to the research community by providing insights that a filter, usually treated as an obstacle, can tell how much a query must read.

\section*{Acknowledgement}
This research was sponsored by the National Research Foundation of Korea (NRF) grants funded by the Ministry of Science and ICT (No. RS-2025-00522990). Gyuyeong Kim is the corresponding author.
\FloatBarrier %
\label{lastpage1}

\bibliographystyle{plain}
\bibliography{./kuroann}

\begin{thebibliography}{10}

\bibitem{darth}
Manos Chatzakis, Yannis Papakonstantinou, and Themis Palpanas.
\newblock {DARTH}: Declarative recall through early termination for approximate
  nearest neighbor search.
\newblock {\em Proceedings of the ACM on Management of Data}, 3(4), 2025.
\newblock Article 242.

\bibitem{bgem3}
Jianlyu Chen, Shitao Xiao, Peitian Zhang, Kun Luo, Defu Lian, and Zheng Liu.
\newblock {M3-Embedding}: Multi-linguality, multi-functionality,
  multi-granularity text embeddings through self-knowledge distillation.
\newblock In {\em Findings of the Association for Computational Linguistics:
  ACL 2024}, pages 2318--2335, 2024.

\bibitem{spann}
Qi~Chen, Bing Zhao, Haidong Wang, Mingqin Li, Chuanjie Liu, Zengzhong Li, Mao
  Yang, and Jingdong Wang.
\newblock {SPANN}: Highly-efficient billion-scale approximate nearest neighbor
  search.
\newblock In {\em Advances in Neural Information Processing Systems (NeurIPS)},
  volume~34, pages 5199--5212, 2021.

\bibitem{chenpref}
Shimin Chen, Anastassia Ailamaki, Phillip~B. Gibbons, and Todd~C. Mowry.
\newblock Improving hash join performance through prefetching.
\newblock {\em ACM Transactions on Database Systems}, 32(3), 2007.
\newblock Article 17.

\bibitem{covington2016youtube}
Paul Covington, Jay Adams, and Emre Sargin.
\newblock Deep neural networks for {YouTube} recommendations.
\newblock In {\em Proceedings of the 10th ACM Conference on Recommender Systems
  (RecSys)}, pages 191--198, 2016.

\bibitem{dade}
Liwei Deng, Penghao Chen, Ximu Zeng, Tianfu Wang, Yan Zhao, and Kai Zheng.
\newblock Efficient data-aware distance comparison operations for
  high-dimensional approximate nearest neighbor search.
\newblock {\em Proceedings of the VLDB Endowment}, 18(3):812--821, 2024.

\bibitem{faiss}
Matthijs Douze, Alexandr Guzhva, Chengqi Deng, Jeff Johnson, Gergely Szilvasy,
  Pierre-Emmanuel Mazar{\'e}, Maria Lomeli, Lucas Hosseini, and Herv{\'e}
  J{\'e}gou.
\newblock The faiss library.
\newblock {\em IEEE Transactions on Big Data}, 12(2):346--361, 2026.

\bibitem{rabitq}
Jianyang Gao and Cheng Long.
\newblock {RaBitQ}: Quantizing high-dimensional vectors with a theoretical
  error bound for approximate nearest neighbor search.
\newblock {\em Proceedings of the ACM on Management of Data}, 2(3), 2024.
\newblock Article 167.

\bibitem{fdiskann}
Siddharth Gollapudi, Neel Karia, Varun Sivashankar, Ravishankar Krishnaswamy,
  Nikit Begwani, Swapnil Raz, Yiyong Lin, Yin Zhang, Neelam Mahapatro,
  Premkumar Srinivasan, Amit Singh, and Harsha~Vardhan Simhadri.
\newblock Filtered-{DiskANN}: Graph algorithms for approximate nearest neighbor
  search with filters.
\newblock In {\em Proceedings of the ACM Web Conference (WWW)}, pages
  3406--3416, 2023.

\bibitem{pipeann}
Hao Guo and Youyou Lu.
\newblock Achieving low-latency graph-based vector search via aligning
  best-first search algorithm with {SSD}.
\newblock In {\em 19th USENIX Symposium on Operating Systems Design and
  Implementation (OSDI)}, pages 171--186, 2025.

\bibitem{odinann}
Hao Guo and Youyou Lu.
\newblock {OdinANN}: Direct insert for consistently stable performance in
  billion-scale graph-based vector search.
\newblock In {\em 24th USENIX Conference on File and Storage Technologies
  (FAST)}, pages 133--147, 2026.

\bibitem{pipeannfilter}
Hao Guo, Jiwu Shu, and Youyou Lu.
\newblock {PipeANN-Filter}: An efficient filtered vector search system on
  {SSD}.
\newblock {\em arXiv preprint arXiv:2605.17992}, 2026.

\bibitem{unwrittenssd}
Jun He, Sudarsun Kannan, Andrea~C. Arpaci-Dusseau, and Remzi~H. Arpaci-Dusseau.
\newblock The unwritten contract of solid state drives.
\newblock In {\em Proceedings of the 12th European Conference on Computer
  Systems (EuroSys)}, pages 127--144, 2017.

\bibitem{blair}
Yupeng Hou, Jiacheng Li, Zhankui He, An~Yan, Xiusi Chen, and Julian McAuley.
\newblock Bridging language and items for retrieval and recommendation.
\newblock {\em arXiv preprint arXiv:2403.03952}, 2024.

\bibitem{hedrarag}
Zhengding Hu, Vibha Murthy, Zaifeng Pan, Wanlu Li, Xiaoyi Fang, Yufei Ding, and
  Yuke Wang.
\newblock {HedraRAG}: Co-optimizing generation and retrieval for heterogeneous
  {RAG} workflows.
\newblock In {\em Proceedings of the 31st Symposium on Operating Systems
  Principles (SOSP)}, pages 623--638, 2025.

\bibitem{huang2020fbsearch}
Jui-Ting Huang, Ashish Sharma, Shuying Sun, Li~Xia, David Zhang, Philip Pronin,
  Janani Padmanabhan, Giuseppe Ottaviano, and Linjun Yang.
\newblock Embedding-based retrieval in {Facebook} search.
\newblock In {\em Proceedings of the 26th ACM SIGKDD International Conference
  on Knowledge Discovery and Data Mining (KDD)}, pages 2553--2561, 2020.

\bibitem{helmsman}
Yuchen Huang, Baiteng Ma, Yiping Sun, Yang Shi, Xiao Chen, Xiaocheng Zhong,
  Zhiyong Wang, Yao Hu, Erci Xu, and Chuliang Weng.
\newblock The clustering strikes back: Building cost-effective and
  high-performance {ANNS} at scale with helmsman.
\newblock In {\em 20th USENIX Symposium on Operating Systems Design and
  Implementation (OSDI)}, pages 1623--1639, 2026.

\bibitem{pinecone}
Amir Ingber and Edo Liberty.
\newblock Accurate and efficient metadata filtering in {Pinecone}'s serverless
  vector database.
\newblock In {\em The 1st Workshop on Vector Databases at {ICML}}, 2025.
\newblock \url{https://openreview.net/forum?id=UXq4z6GGYP}.

\bibitem{pqo}
Yannis~E. Ioannidis, Raymond~T. Ng, Kyuseok Shim, and Timos~K. Sellis.
\newblock Parametric query optimization.
\newblock In {\em Proceedings of the 18th International Conference on Very
  Large Data Bases}, pages 103--114, 1992.

\bibitem{pq}
Herv{\'e} J{\'e}gou, Matthijs Douze, and Cordelia Schmid.
\newblock Product quantization for nearest neighbor search.
\newblock {\em IEEE Transactions on Pattern Analysis and Machine Intelligence},
  33(1):117--128, 2011.

\bibitem{bigann11}
Herv{\'e} J{\'e}gou, Romain Tavenard, Matthijs Douze, and Laurent Amsaleg.
\newblock Searching in one billion vectors: Re-rank with source coding.
\newblock In {\em IEEE International Conference on Acoustics, Speech and Signal
  Processing (ICASSP)}, pages 861--864, 2011.

\bibitem{curator26}
Yicheng Jin, Yongji Wu, Wenjun Hu, Bruce~M. Maggs, Jun Yang, Xiao Zhang, and
  Danyang Zhuo.
\newblock Curator: Efficient vector search with low-selectivity filters.
\newblock {\em Proceedings of the ACM on Management of Data}, 4(1), 2026.
\newblock Article 21.

\bibitem{iopassthru}
Kanchan Joshi, Anuj Gupta, Javier Gonz{\'a}lez, Ankit Kumar, Krishna~Kanth
  Reddy, Arun George, Simon Lund, and Jens Axboe.
\newblock {I/O} passthru: Upstreaming a flexible and efficient {I/O} path in
  {Linux}.
\newblock In {\em 22nd USENIX Conference on File and Storage Technologies
  (FAST)}, pages 107--121, 2024.

\bibitem{ingens}
Youngjin Kwon, Hangchen Yu, Simon Peter, Christopher~J. Rossbach, and Emmett
  Witchel.
\newblock Coordinated and efficient huge page management with {Ingens}.
\newblock In {\em 12th USENIX Symposium on Operating Systems Design and
  Implementation (OSDI)}, pages 705--721, 2016.

\bibitem{ivf2}
Ben Landrum, Magdalen Dobson~Manohar, Mazin Karjikar, and Laxman Dhulipala.
\newblock {IVF}$^{2}$ index: Fusing classic and spatial inverted indices for
  fast filtered {ANNS}.
\newblock In {\em The 1st Workshop on Vector Databases at {ICML}}, 2025.
\newblock \url{https://openreview.net/forum?id=kXw8E3xT7O}.

\bibitem{nemoann}
Nakyung Lee, Soobin Cho, Jiwoong Park, and Gyuyeong Kim.
\newblock {NemoANN}: {I/O}-efficient filtered vector search on {SSDs} with
  graph tunneling.
\newblock In {\em ACM Symposium on Cloud Computing (SoCC)}, 2026.

\bibitem{lewis2020rag}
Patrick Lewis, Ethan Perez, Aleksandra Piktus, Fabio Petroni, Vladimir
  Karpukhin, Naman Goyal, Heinrich K{\"u}ttler, Mike Lewis, Wen-tau Yih, Tim
  Rockt{\"a}schel, Sebastian Riedel, and Douwe Kiela.
\newblock Retrieval-augmented generation for knowledge-intensive {NLP} tasks.
\newblock In {\em Advances in Neural Information Processing Systems (NeurIPS)},
  volume~33, pages 9459--9474, 2020.

\bibitem{earlyterm}
Conglong Li, Minjia Zhang, David~G. Andersen, and Yuxiong He.
\newblock Improving approximate nearest neighbor search through learned
  adaptive early termination.
\newblock In {\em Proceedings of the 2020 ACM SIGMOD International Conference
  on Management of Data}, pages 2539--2554, 2020.

\bibitem{fannbench}
Mocheng Li, Xiao Yan, Baotong Lu, Yue Zhang, James Cheng, and Chenhao Ma.
\newblock Attribute filtering in approximate nearest neighbor search: An
  in-depth experimental study.
\newblock {\em Proceedings of the ACM on Management of Data}, 3(6), 2025.
\newblock Article 298.

\bibitem{sieve}
Zhaoheng Li, Silu Huang, Wei Ding, Yongjoo Park, and Jianjun Chen.
\newblock {SIEVE}: Effective filtered vector search with collection of indexes.
\newblock {\em Proceedings of the VLDB Endowment}, 18(11):4723--4736, 2025.

\bibitem{pathseer}
Zhiyue Li, Guangyan Zhang, Ruochun Jin, Bojun Li, Xiaoguang Ren, and Wenjing
  Yang.
\newblock {PathSeer}: Adaptive neighbor handling for efficient filtered {ANN}
  search.
\newblock {\em Proceedings of the ACM on Management of Data}, 4(3), 2026.
\newblock Article 221.

\bibitem{unify}
Anqi Liang, Pengcheng Zhang, Bin Yao, Zhongpu Chen, Yitong Song, and Guangxu
  Cheng.
\newblock {UNIFY}: Unified index for range filtered approximate nearest
  neighbors search.
\newblock {\em Proceedings of the VLDB Endowment}, 18(4):1118--1130, 2024.

\bibitem{lufvs26}
Duo Lu, Helena Caminal, Manos Chatzakis, Yannis Papakonstantinou, Yannis
  Chronis, Vaibhav Jain, and Fatma {\"O}zcan.
\newblock An in-depth study of filter-agnostic vector search on a {PostgreSQL}
  database system.
\newblock {\em Proceedings of the ACM on Management of Data}, 4(3), 2026.
\newblock Article 134.

\bibitem{tagfilter}
Jiarui Luo, Miao Qiao, Chaoji Zuo, and Dong Deng.
\newblock Tag-filtered approximate nearest neighbor search.
\newblock In {\em Proceedings of the IEEE International Conference on Data
  Engineering (ICDE)}, pages 3642--3654, 2025.

\bibitem{hnsw}
Yu~A. Malkov and Dmitry~A. Yashunin.
\newblock Efficient and robust approximate nearest neighbor search using
  hierarchical navigable small world graphs.
\newblock {\em IEEE Transactions on Pattern Analysis and Machine Intelligence},
  42(4):824--836, 2020.

\bibitem{quake}
Jason Mohoney, Devesh Sarda, Mengze Tang, Shihabur~Rahman Chowdhury, Anil
  Pacaci, Ihab~F. Ilyas, Theodoros Rekatsinas, and Shivaram Venkataraman.
\newblock Quake: Adaptive indexing for vector search.
\newblock In {\em 19th USENIX Symposium on Operating Systems Design and
  Implementation (OSDI 25)}, pages 153--169, 2025.

\bibitem{acorn}
Liana Patel, Peter Kraft, Carlos Guestrin, and Matei Zaharia.
\newblock {ACORN}: Performant and predicate-agnostic search over vector
  embeddings and structured data.
\newblock {\em Proceedings of the ACM on Management of Data}, 2(3), 2024.
\newblock Article 120.

\bibitem{micronn}
Jeffrey Pound, Floris Chabert, Arjun Bhushan, Ankur Goswami, Anil Pacaci, and
  Shihabur~Rahman Chowdhury.
\newblock {MicroNN}: An on-device disk-resident updatable vector database.
\newblock In {\em Companion of the 2025 International Conference on Management
  of Data (SIGMOD-Companion)}, pages 608--621, 2025.

\bibitem{qdrantplanning}
{Qdrant}.
\newblock Query planning, {Qdrant} documentation.
\newblock
  \url{https://qdrant.tech/documentation/search/search/\#query-planning}.
\newblock Accessed November 2, 2026.

\bibitem{pubmedmv}
{Qdrant}.
\newblock {PubMed-MultiVector}.
\newblock \url{https://huggingface.co/datasets/Qdrant/PubMed-MV}, 2026.

\bibitem{iks}
Derrick Quinn, Mohammad Nouri, Neel Patel, John Salihu, Alireza Salemi, Sukhan
  Lee, Hamed Zamani, and Mohammad Alian.
\newblock Accelerating retrieval-augmented generation.
\newblock In {\em Proceedings of the 30th ACM International Conference on
  Architectural Support for Programming Languages and Operating Systems
  (ASPLOS)}, pages 15--32, 2025.

\bibitem{openclosed}
Bianca Schroeder, Adam Wierman, and Mor Harchol-Balter.
\newblock Open versus closed: A cautionary tale.
\newblock In {\em 3rd USENIX Symposium on Networked Systems Design and
  Implementation (NSDI)}, pages 239--252, 2006.

\bibitem{laion5b}
Christoph Schuhmann, Romain Beaumont, Richard Vencu, Cade Gordon, Ross
  Wightman, Mehdi Cherti, Theo Coombes, Aarush Katta, Clayton Mullis, Mitchell
  Wortsman, Patrick Schramowski, Srivatsa Kundurthy, Katherine Crowson, Ludwig
  Schmidt, Robert Kaczmarczyk, and Jenia Jitsev.
\newblock {LAION-5B}: An open large-scale dataset for training next generation
  image-text models.
\newblock In {\em Advances in Neural Information Processing Systems (NeurIPS),
  Datasets and Benchmarks Track}, volume~35, pages 25278--25294, 2022.

\bibitem{multistepknn}
Thomas Seidl and Hans-Peter Kriegel.
\newblock Optimal multi-step k-nearest neighbor search.
\newblock In {\em Proceedings of the 1998 ACM SIGMOD International Conference
  on Management of Data}, pages 154--165, 1998.

\bibitem{trim}
Yitong Song, Pengcheng Zhang, Chao Gao, Bin Yao, Kai Wang, Zongyuan Wu, and Lin
  Qu.
\newblock {TRIM}: Accelerating high-dimensional vector similarity search with
  enhanced triangle-inequality-based pruning.
\newblock {\em Proceedings of the ACM on Management of Data}, 3(6), 2025.
\newblock Article 373.

\bibitem{diskann}
Suhas~Jayaram Subramanya, Devvrit, Rohan Kadekodi, Ravishankar Krishnaswamy,
  and Harsha~Vardhan Simhadri.
\newblock {DiskANN}: Fast accurate billion-point nearest neighbor search on a
  single node.
\newblock In {\em Advances in Neural Information Processing Systems (NeurIPS)},
  volume~32, 2019.

\bibitem{smartanns}
Bing Tian, Haikun Liu, Zhuohui Duan, Xiaofei Liao, Hai Jin, and Yu~Zhang.
\newblock Scalable billion-point approximate nearest neighbor search using
  {SmartSSDs}.
\newblock In {\em 2024 USENIX Annual Technical Conference (USENIX ATC)}, pages
  1135--1150, 2024.

\bibitem{fusionanns}
Bing Tian, Haikun Liu, Yuhang Tang, Shihai Xiao, Zhuohui Duan, Xiaofei Liao,
  Hai Jin, Xuecang Zhang, Junhua Zhu, and Yu~Zhang.
\newblock Towards high-throughput and low-latency billion-scale vector search
  via {CPU/GPU} collaborative filtering and re-ranking.
\newblock In {\em 23rd USENIX Conference on File and Storage Technologies (FAST
  25)}, 2025.

\bibitem{ubase}
Yao Tian, Yuncheng Lu, Liyao Xiong, Yuming Xu, Hao Zhang, Weichen Zhao,
  Xi~Zhao, Bo~Kuang, Dongyu Wang, Jiehui Li, Yakun Li, and Lei Zhang.
\newblock {UBASE}: An {AI} search engine for trillion-scale vector data
  management at {ByteDance}.
\newblock In {\em ACM SIGOPS Annual Technical Conference (ATC)}, 2026.

\bibitem{cosmosdb}
Nitish Upreti, Harsha~Vardhan Simhadri, Hari~Sudan Sundar, Krishnan Sundaram,
  Samer Boshra, et~al.
\newblock Cost-effective, low latency vector search with {Azure Cosmos DB}.
\newblock {\em Proceedings of the VLDB Endowment}, 18(12):5166--5183, 2025.

\bibitem{vaidya2021pqo}
Kapil Vaidya, Anshuman Dutt, Vivek Narasayya, and Surajit Chaudhuri.
\newblock Leveraging query logs and machine learning for parametric query
  optimization.
\newblock {\em Proceedings of the VLDB Endowment}, 15(3):401--413, 2021.

\bibitem{milvus}
Jianguo Wang, Xiaomeng Yi, Rentong Guo, Hai Jin, Peng Xu, Shengjun Li, Xiangyu
  Wang, Xiangzhou Guo, Chengming Li, Xiaohai Xu, et~al.
\newblock Milvus: A purpose-built vector data management system.
\newblock In {\em Proceedings of the 2021 International Conference on
  Management of Data (SIGMOD)}, pages 2614--2627, 2021.

\bibitem{nhq}
Mengzhao Wang, Lingwei Lv, Xiaoliang Xu, Yuxiang Wang, Qiang Yue, and Jiongkang
  Ni.
\newblock An efficient and robust framework for approximate nearest neighbor
  search with attribute constraint.
\newblock In {\em Advances in Neural Information Processing Systems (NeurIPS)},
  volume~36, pages 15738--15751, 2023.

\bibitem{starling}
Mengzhao Wang, Weizhi Xu, Xiaomeng Yi, Songlin Wu, Zhangyang Peng, Xiangyu Ke,
  Yunjun Gao, Xiaoliang Xu, Rentong Guo, and Charles Xie.
\newblock Starling: An {I/O}-efficient disk-resident graph index framework for
  high-dimensional vector similarity search on data segment.
\newblock {\em Proceedings of the ACM on Management of Data}, 2(1), 2024.
\newblock Article 14.

\bibitem{weaviatefiltering}
{Weaviate}.
\newblock Filtering, {Weaviate} documentation.
\newblock \url{https://docs.weaviate.io/weaviate/concepts/filtering}.
\newblock Accessed November 2, 2026.

\bibitem{adbv}
Chuangxian Wei, Bin Wu, Sheng Wang, Renjie Lou, Chaoqun Zhan, Feifei Li, and
  Yuanzhe Cai.
\newblock {AnalyticDB-V}: A hybrid analytical engine towards query fusion for
  structured and unstructured data.
\newblock {\em Proceedings of the VLDB Endowment}, 13(12):3152--3165, 2020.

\bibitem{vecflow}
Jingyi Xi, Chenghao Mo, Ben Karsin, Artem Chirkin, Mingqin Li, and Minjia
  Zhang.
\newblock {VecFlow}: A high-performance vector data management system for
  filtered-search on {GPUs}.
\newblock {\em Proceedings of the ACM on Management of Data}, 3(4), 2025.
\newblock Article 271.

\bibitem{e2e}
Wenxuan Xia, Mingyu Yang, Wentao Li, and Wei Wang.
\newblock {E2E}: Efficient filtered {AKNN} search via adaptive termination.
\newblock In {\em Proceedings of the 32nd ACM SIGKDD Conference on Knowledge
  Discovery and Data Mining V.2 (KDD)}, pages 5602--5613, 2026.

\bibitem{irangegraph}
Yuexuan Xu, Jianyang Gao, Yutong Gou, Cheng Long, and Christian~S. Jensen.
\newblock {iRangeGraph}: Improvising range-dedicated graphs for range-filtering
  nearest neighbor search.
\newblock {\em Proceedings of the ACM on Management of Data}, 2(6), 2024.
\newblock Article 239.

\bibitem{rabitqssd}
Yuexuan Xu, Jianyang Gao, Michael Norris, Alibek Zhakubayev, Pankaj Singh,
  Junjie Qi, Matthijs Douze, and Cheng Long.
\newblock {RaBitQ-SSD}: Split codes and pipelined {I/O} for {SSD}-resident
  vector search.
\newblock {\em arXiv preprint arXiv:2610.02652}, 2026.

\bibitem{spfresh}
Yuming Xu, Hengyu Liang, Jin Li, Shuotao Xu, Qi~Chen, Qianxi Zhang, Cheng Li,
  Ziyue Yang, Fan Yang, Yuqing Yang, Peng Cheng, and Mao Yang.
\newblock {SPFresh}: Incremental in-place update for billion-scale vector
  search.
\newblock In {\em Proceedings of the 29th Symposium on Operating Systems
  Principles (SOSP)}, pages 545--561, 2023.

\bibitem{distvs}
Peiqi Yin, Xiao Yan, Shiyuan Deng, Hui Li, Yifan Zhu, Xiangyu Zhi, Jingqi Mao,
  Ran Xu, Wenliang Zhang, and James Cheng.
\newblock {DistVS}: Large-scale vector search with compute-memory
  disaggregation.
\newblock In {\em 23rd USENIX Symposium on Networked Systems Design and
  Implementation (NSDI)}, pages 449--467, 2026.

\bibitem{adaef}
Chao Zhang and Ren{\'e}e~J. Miller.
\newblock Distribution-aware exploration for adaptive {HNSW} search.
\newblock {\em Proceedings of the ACM on Management of Data}, 2026.

\bibitem{zoom}
Minjia Zhang and Yuxiong He.
\newblock Zoom: {SSD}-based vector search for optimizing accuracy, latency and
  memory.
\newblock {\em arXiv preprint arXiv:1809.04067}, 2018.

\bibitem{grip}
Minjia Zhang and Yuxiong He.
\newblock {GRIP}: Multi-store capacity-optimized high-performance nearest
  neighbor search for vector search engine.
\newblock In {\em Proceedings of the 28th ACM International Conference on
  Information and Knowledge Management (CIKM)}, pages 1673--1682, 2019.

\bibitem{vbase}
Qianxi Zhang, Shuotao Xu, Qi~Chen, Guoxin Sui, Jiadong Xie, Zhizhen Cai, Yaoqi
  Chen, Yinxuan He, Yuqing Yang, Fan Yang, Mao Yang, and Lidong Zhou.
\newblock {VBASE}: Unifying online vector similarity search and relational
  queries via relaxed monotonicity.
\newblock In {\em 17th USENIX Symposium on Operating Systems Design and
  Implementation (OSDI)}, pages 377--395, 2023.

\bibitem{auncel}
Zili Zhang, Chao Jin, Linpeng Tang, Xuanzhe Liu, and Xin Jin.
\newblock Fast, approximate vector queries on very large unstructured datasets.
\newblock In {\em 20th USENIX Symposium on Networked Systems Design and
  Implementation (NSDI 23)}, pages 995--1011, 2023.

\bibitem{rummy}
Zili Zhang, Fangyue Liu, Gang Huang, Xuanzhe Liu, and Xin Jin.
\newblock Fast vector query processing for large datasets beyond {GPU} memory
  with reordered pipelining.
\newblock In {\em 21st USENIX Symposium on Networked Systems Design and
  Implementation (NSDI)}, pages 23--40, 2024.

\bibitem{flowann}
Haoru Zhao, Jingkai He, Jingyao Zeng, Mingkai Dong, and Dong Du.
\newblock Disentangling graph dependencies for efficient billion-scale {GPU}
  vector search.
\newblock In {\em 20th USENIX Symposium on Operating Systems Design and
  Implementation (OSDI)}, pages 1563--1584, 2026.

\bibitem{flashgraph}
Da~Zheng, Disa Mhembere, Randal Burns, Joshua Vogelstein, Carey~E. Priebe, and
  Alexander~S. Szalay.
\newblock {FlashGraph}: Processing billion-node graphs on an array of commodity
  {SSDs}.
\newblock In {\em 13th USENIX Conference on File and Storage Technologies
  (FAST)}, pages 45--58, 2015.

\bibitem{xrp}
Yuhong Zhong, Haoyu Li, Yu~Jian Wu, Ioannis Zarkadas, Jeffrey Tao, Evan
  Mesterhazy, Michael Makris, Junfeng Yang, Amy Tai, Ryan Stutsman, and Asaf
  Cidon.
\newblock {XRP}: {In-Kernel} storage functions with {eBPF}.
\newblock In {\em 16th USENIX Symposium on Operating Systems Design and
  Implementation (OSDI)}, pages 375--393, 2022.

\bibitem{vdbb}
{Zilliz}.
\newblock {VectorDBBench (VDBBench)}: A benchmark tool for {VectorDB}.
\newblock \url{https://github.com/zilliztech/VectorDBBench}, 2023.

\bibitem{serf}
Chaoji Zuo, Miao Qiao, Wenchao Zhou, Feifei Li, and Dong Deng.
\newblock {SeRF}: Segment graph for range-filtering approximate nearest
  neighbor search.
\newblock {\em Proceedings of the ACM on Management of Data}, 2(1), 2024.
\newblock Article 69.

\end{thebibliography}

\label{lastpage2}

\end{document}